\documentclass[useAMS,usenatbib]{mn2e}
\usepackage[dvips]{graphicx}
\usepackage{amssymb}
\usepackage{longtable}
\newcommand{\dn}{\hbox{$\rm D4000$}}
\newcommand{\hb}{\hbox{H$\beta$}}
\newcommand{\hdg}{\hbox{H$\delta_A$+H$\gamma_A$}}
\newcommand{\mgfep}{\hbox{$\rm [MgFe]^\prime$}}

\newcommand{\mgtwofe}{\hbox{$\rm [Mg_2Fe]$}}

\newcommand{\sigv}{\hbox{$\sigma_V$}}

\newcommand{\vmax}{\hbox{$\rm V_{max}$}}

\def\aj{AJ}%
\def\araa{ARA\&A}%
\def\apj{ApJ}%
\def\apjl{ApJ}%
\def\apjs{ApJS}%
\def\aap{A\&A}%
\def\mnras{MNRAS}%
\def\pasp{PASP}%
\def\nat{Nature}%
\title[A local census of metals and baryons in stars]
{A census of metals and baryons in stars in the local Universe}
\author[A. Gallazzi, J. Brinchmann, S. Charlot, S.D.M. White]{Anna Gallazzi$^{1,2}$\thanks{E-mail:
gallazzi@mpia-hd.mpg.de}, Jarle Brinchmann$^{3}$, St{\'e}phane Charlot$^{2,4}$, Simon D.M. White$^{2}$\\
$^{1}$Max-Planck-Institut f\"ur Astronomie, K\"onigstuhl 17, 
D-69117 Heidelberg, Germany\\
$^{2}$Max-Planck-Institut f\"ur Astrophysik, Karl-Schwarzschild-Str. 1, 
D-85748 Garching bei M\"unchen, Germany\\
$^{3}$Centro de Astrof{\'\i}sica da
Universidade do Porto, Rua das Estrelas - 4150-762 Porto, Portugal\\
$^{4}$Institut d'Astrophysique de Paris, UMR7095 CNRS, Universit\'e Pierre \& Marie Curie, 98 bis
boulevard Arago, 75014 Paris, France}
\begin{document}

\date{Accepted 2007 October 23. Received 2007 October 23; in original form 2007 August 3}

\pagerange{\pageref{firstpage}--\pageref{lastpage}} \pubyear{2007}

\maketitle

  \label{firstpage}

\begin{abstract}
We combine stellar metallicity and stellar mass estimates for a large sample of galaxies drawn from the
Sloan Digital Sky Survey Data Release Two (SDSS DR2) spanning wide ranges in physical properties, in
order to derive an inventory of the total mass of metals and baryons locked up in stars in the local
Universe. Physical parameter estimates are derived from galaxy spectra with high signal-to-noise (S/N)
ratio (of at least 20). Coadded spectra of galaxies with similar velocity dispersions, absolute $r$-band
magnitudes and 4000\AA-break values are used for those regions of parameter space where individual
spectra have lower S/N. We estimate the total density of metals $\rho_Z$ and of baryons $\rho_\ast$ in
stars and, from these two quantities, we obtain a mass- and volume-averaged stellar metallicity of
$\langle Z_\ast\rangle=1.04\pm0.14~Z_\odot$, i.e.  consistent with solar. We also study how metals are
distributed in galaxies according to different properties, such as mass, morphology, mass- and
light-weighted age, and we then compare these distributions with the corresponding distributions of
stellar mass. We find that the bulk of metals locked up in stars in the local Universe reside in massive,
bulge-dominated galaxies, with red colours and high 4000\AA-break values corresponding to old stellar
populations. Bulge-dominated and disc-dominated galaxies contribute similar amounts to the total stellar
mass density, but have different fractional contributions to the mass density of metals in stars, in
agreement with the mass-metallicity relation. Bulge-dominated galaxies contain roughly 40 percent of the
total amount of metals in stars, while disc-dominated galaxies less than 25 percent. Finally, at a given
galaxy stellar mass, we define two characteristic ages as the median of the distributions of mass and
metals as a function of age. These characteristic ages decrease progressively from high-mass to low-mass
galaxies, consistent with the high formation epochs of stars in massive galaxies.
\end{abstract}


\section{Introduction}
Constraining the star formation and chemical evolution histories of galaxies is one of the fundamental
goals in observational cosmology. The evolution of the global star formation rate (SFR) has to map the
evolution over cosmic time of its products, i.e. the baryonic and metal content of the Universe.  

The most direct way to constrain the star formation and chemical evolution history over cosmic times is
to trace back galaxy properties (star formation rate, metallicity, stellar mass) through observations at
different redshifts. Several studies on the evolution of the rest-frame UV emission density of galaxies,
converted into star formation or metal ejection rates, have converged into a picture in which the maximum
of galaxy star formation activity occurs over the redshift range $1\la z\la 2$ and declines sharply from
$z=1$ towards the present \citep{lilly96,connolly97,cowie99}. While several recent studies have built a
consistent picture of the decline in cosmic star formation rate from $z\sim$1 to the present
\citep[e.g.][and references therein]{hopkins06}, more uncertain is the behaviour at redshift higher than
2, because of the poor understanding of the effect of dust on the SFR derived from UV Spectral Energy
Distributions (SED) of high-redshift galaxies \citep{madau96,steidel99,ivison02}. A broad peak of high
star formation over the redshift range $z\sim1-2$ and then a rapid decline toward the present are
features predicted (or reproduced) also by both chemical evolution models \citep{pei95,edmunds97,pei99}
and by semianalytic models of galaxy formation \citep[e.g.][]{baugh98}.

A complementary approach is to study the chemical and star formation history over cosmic times through
the so-called `fossil cosmology', i.e. determining the past history of the Universe from its present
contents. This approach has benefited from large spectroscopic surveys in the local Universe, such as the
2dF Galaxy Redshift Survey \citep[2dFGRS,][]{colless01} and the Sloan Digital Sky Survey
\citep[SDSS,][]{York}, which provide detailed spectral information for hundreds of thousands of galaxies.
Based on such surveys, \cite{baldry02} and \cite{glazebrook03} have constrained the cosmic star formation
history (SFH) from the `cosmic optical spectrum', which represents the average emission from all the
objects in a representative volume of the Universe and has the advantage of being fitted by simpler
models of star formation histories than those needed for individual objects.  \cite{heavens04} and
\cite{jimenez05} have applied a data compression algorithm \citep[MOPED,][]{moped} to extract the SFH of
$\sim100,000$ SDSS DR1 galaxies from their optical spectra. This work has been recently extended to the
SDSS DR3 (three times larger sample) by \cite{ben06}, with improvements both in the data and in the
modelling sides (see also Tojeiro et al. 2007, Ocvirk et al. 2006 and Cid Fernandes et al. 2007 for similar methods to recover
stellar content and star formation histories from galaxy spectra). This allowed them to derive the cosmic SFH from the `fossil record' and study it as a
function of the present-day stellar mass of galaxies. 

The contribution to the global SFR by galaxies of different mass is being studied not only in the local
Universe but also at higher redshift. \cite{juneau05} studied the dependence of the cosmic SFH 
directly on the stellar mass at the epoch of observation over the redshift range $0.8<z<2$, based
on a near-infrared selected sample from the Gemini Deep Deep Survey \citep[GDDS,][]{GDDS}. Similarly,
\cite{bundy06} have quantified the decrease with redshift of the mass limit above which star formation
appears to be quenched, based on a sample of more than 8000 galaxies in the redshift range $0.4<z<1.4$
drawn from the DEEP2 Galaxy Redshift Survey \citep{deep2}. These results confirm those previously found
by \cite{jarle2000} for $0<z<1$. While high- and intermediate-mass galaxies have transitioned to a
quiescent phase of star formation by $z\sim1$, less massive systems dominate the star formation rate
density till the present epoch. It appears, though, that the global star formation rate has been
declining since $z\sim$1 for all galaxies populations, at a rate which is independent of stellar mass, as
shown by \cite{xxz07} estimating SFRs of $\sim$15000 COMBO-17 galaxies from UV and IR luminosities and
accounting for individually IR-undetectable galaxies \citep{xxz06}.

An important consistency check for all these studies comes from the comparison of the density of stellar
mass and of metals at different epochs expected from the cosmic star formation and chemical enrichment
histories  (i.e. the integral of these histories) with those directly measured. The evolution of the
global stellar mass density out to $z=3$ has been first determined by \cite{dickinson03}. In concordance
with estimates of the cosmic SFH, their study suggests that the redshift range $1<z<2.5$ is a critical
epoch when galaxies are growing rapidly attaining their final stellar mass. 

Much effort has been put also in measuring the chemical composition of galaxies at different epochs,
through optical nebular emission lines studies at $z<1$ \citep[e.g.][]{kz99,lilly03,kk04,ellison05},
through Lyman-break and UV-selected star-forming galaxies up to $z\sim3$
\citep[e.g.][]{pettini01,steidel04,shapley04,erb06}, through quasar absorption-line systems, in
particular Damped Ly-$\alpha$ Absorbers (DLA) at any redshift
\citep[e.g.][]{pettini94,lanzetta95,pettini97,peroux03,peroux05,peroux06}. All these studies have
highlighted a shortfall of metals in observed galaxy populations with respect to expectations from the
cosmic SFH, known as the `missing metals' problem at redshift around 2.5 \citep{pettini06,bouche05}. A
large fraction could be hosted in a recently discovered population of sub-DLAs \citep{peroux05}. Not more
than $30-40$ percent seems to be in intergalactic medium (IGM), and probably $\la 35$ percent of metals
is still `missing' from the census \citep{bouche07}. These metals are likely locked in the hot gas phase
\citep{ferrara05,DO07}. The present-day distribution of metals is still highly uncertain, because little
is known about the chemical composition of the possibly dominant baryonic component, the warm hot
intergalactic medium (WHIM). However the fraction of metals contained in galaxies, in particular those
locked into the stellar component, has increased from $z\sim2$ to the present, and is probably comparable
to the fraction of metals outside galaxies \citep[e.g.][]{dunne03,cm2004}. 

In this work we focus on the baryonic and metal content of the stellar component of galaxies in the local
Universe. What is the total amount of metals and baryons locked up into stars by the present epoch? What
is the resulting average stellar metallicity of the present-day Universe? To address these questions we
join together information about the stellar mass and chemical properties of present-day galaxies,
supported by the large statistics provided by the SDSS. The sample we analyse span large ranges in
physical, spectral and morphological properties, and constitute in this sense a representative sample of
the local Universe. This allows us also to study how metals and baryons are distributed among galaxies with
different properties. In particular, we want to quantify the fraction of metals, in comparison to the
fraction of baryons, locked up in galaxies as a function of their stellar mass, morphology and age.

We exploit new estimates of physical parameters, such as stellar metallicity and stellar mass, that we
previously derived \citep[][hereafter paper~I]{paperI} for a large sample of nearly $2\times10^5$
galaxies drawn from the Sloan Digital Sky Survey Data Release Two (SDSS DR2). We include all galaxies
types, from quiescent early-type to actively star-forming galaxies. In our previous works we focused only
on galaxies with high signal-to-noise ratio (S/N) spectra, because of the poor constraints that can be obtained
on stellar metallicity from low-S/N spectra. We circumvent here this problem by stacking individual
spectra of low-S/N galaxies with similar properties in order to obtain high-S/N (average) spectra. 

The sample analysed is described in Section~\ref{sample}, along with the stacking technique adopted in
order to include galaxies with low-S/N spectra (Section~\ref{stacking}). We illustrate the physical
parameters estimates extracted from individual galaxy spectra and from coadded spectra in
Section~\ref{estimates}. In Section~\ref{totalz} we derive the mass density of baryons and of metals
locked up in stars, expressed also in terms of the average stellar metallicity of the local Universe. We
then discuss several sources of systematic uncertainties in Section~\ref{errors} and compare with other
estimates in the literature in Section~\ref{literature}. Section~\ref{distrz} provides an inventory of
the stellar metallicity and stellar mass today, focusing on the characteristic age of the stellar mass
and metallicity distributions in Section~\ref{age}. We compare the observed distributions of stellar mass
and stellar metallicity with those predicted by the Millennium Simulation in Section~\ref{millennium}. We
finally summarise and conclude in Section~\ref{conclusions}. Throughout the paper we adopt a flat
cosmology with $\rm \Omega_m=0.3$, $\rm \Omega_\Lambda=0.7$ and $H_0=\rm 70~h_{70}~km~s^{-1}~Mpc^{-1}$.
The models used for this work are computed for a \cite{chabrier03} initial mass function (IMF) and are
based on a metallicity scale where the solar metallicity is $\rm Z_\odot=0.02$. All the magnitudes used
in this work are SDSS model magnitudes, unless otherwise specified. 

\section{The approach}\label{approach}
In this section we give a brief overview of the sample analysed and the method applied to derive
estimates of physical parameters, such as stellar metallicity, (light- and mass-weighted) age and stellar
mass (Section~\ref{sample}, the reader is referred to paper~I for a more thorough description of the
method). The method requires spectra with high S/N. To derive a fair estimate of the total budget of mass
and metals in stars today we need however to include all objects. We include low-S/N galaxies by adopting
a stacking technique, which is described in Section~\ref{stacking}. We compare the physical parameters of
the coadded spectra to those of individual galaxies in Section~\ref{estimates}. 

\subsection{The sample}\label{sample}
To derive an estimate of the total amount of metals and baryons locked up in stars today and to study
their distribution as a function of various galaxy properties we exploit a large sample of galaxies, for
which stellar metallicities, as well as other physical parameters, have been estimated. The sample
analysed here is drawn from the main spectroscopic sample of the SDSS DR2 \citep{dr204} and is based on
164,746 unique spectra of galaxies with Petrosian $r$-band magnitudes in the range $14.5\le r \le 17.77$
\citep[after correction for Galactic extinction using the extinction maps of][]{Schlegel}, and with
redshift\footnote{As explained in paper~I, we choose to limit the analysis in this redshift range, in
order to avoid redshifts for which deviations from the Hubble flow can be substantial and to include
galaxies in the stellar mass range $10^8-10^{11} M_\odot$ with a signal-to-noise per pixel of at least
20.} between 0.005 and 0.22. The sample includes all galaxy types, from star-forming late-type to
quiescent early-type galaxies. We note that the sample analysed is defined on the DR2 coverage, but we
use the photometric reduction of the DR4 release. This is motivated by the fact that we found a
systematic difference of $\sim$0.16~mag in the $z$-band model magnitudes for a subset of galaxies from
one release to the other, which can affect the overall normalization of the stellar mass density. The
spectroscopic measurements and fibre colours were instead consistent within the errors between
releases.\footnote{The stellar metallicity, light-weighted age and stellar mass estimates for the whole
DR4 are available at http://www.mpa-garching.mpg.de/SDSS/DR4/.}

Bayesian-likelihood estimates of the stellar metallicities, $r$-band light-weighted ages and stellar
masses of the galaxies in the sample have been obtained in our previous work, by comparing the spectrum
of each galaxy to a library of \citet[][hereafter BC03]{bc03} models, covering the full range of
physically plausible star formation histories. The comparison is based on five spectral absorption
features, namely \dn, \hb\ and \hdg\ as age-sensitive indices, and \mgtwofe\ and \mgfep\ as
metal-sensitive indices, all of which have at most a weak dependence on element abundance ratios.  After
constructing the probability density function of age, metallicity and stellar mass for every galaxy, the
median of each likelihood  distribution represents our estimate of the corresponding parameter, while
half of the $16-84$ percent interpercentile range gives the associated $\pm 1\sigma$
(Gaussian-equivalent) uncertainty. In this
work we add information about the mass-weighted age of galaxies. In Section~\ref{millennium} we shall use
this quantity also in comparison with predictions from the Millennium Simulation \citep{millennium}. 
We have derived mass-weighted ages in the same way as the other physical parameters as described in
paper~I. The mass-weighted age of each model in the library has been estimated by weighting each
generation of stars by their mass, taking into account the fraction of mass returned to the interstellar
medium (ISM) by long-lived stars.

In \citet[][hereafter paper~II]{paperI,paperII} we focused only on galaxy spectra with median S/N per
pixel of at least 20. As explained there, this is the minimum S/N required in order to obtain reliable
estimates of stellar metallicity. The quality of the spectrum influences directly the uncertainties in
the derived physical parameters, stellar metallicity being the most affected one: the average uncertainty on
stellar metallicity decreases from 0.21~dex to 0.12~dex when  high-S/N galaxies only are considered. The
cut in S/N excludes roughly 75 percent of the galaxies and biases the sample towards high-surface
brightness, high-concentration, low-redshift galaxies. Only 10 percent of the galaxies with concentration
parameter\footnote{defined as the ratio between the radii including 90 and 50 percent of the $r$-band
Petrosian flux.} $C\leq2.4$ satisfies the S/N requirement.  Excluding galaxies with $\rm S/N<20$ we would
therefore preferentially miss diffuse systems with potentially subsolar metallicity. In order to derive a
fair estimate of the total metal budget in the local Universe we need to include all galaxies
down to the magnitude limit of the survey, therefore low-S/N galaxies need to be considered as well.

\subsection{The stacking technique}\label{stacking}
In order to include low-S/N galaxies, in addition to the subsample with $\rm S/N\geq20$, we create
composite high-S/N spectra by coadding the spectra of low-S/N galaxies with similar properties. First of
all, we require galaxies to have similar velocity dispersion. The broadening due to stellar velocity
dispersion affects the measured spectral absorption indices. When deriving physical parameters estimates
we do not correct for this, instead each spectrum is compared only to those models in the library with
velocity dispersion similar to the observed one. It is therefore important that the galaxies that
contribute to each coadded spectrum span a range in velocity dispersion comparable to the observational
error. Moreover, metallicity, age and stellar mass all show correlations with velocity dispersion,
absolute magnitude and \dn\ (see e.g. figs~7,8 of paper~I and figs~6,10 of paper~II for early-type
galaxies). We thus choose to coadd the spectra of low-S/N galaxies with similar velocity dispersion,
$r$-band absolute magnitude and 4000\AA-break. By binning into these quantities we are confident that the
scatter in the physical parameters of the galaxies contributing to each stacked spectrum is small.

We first divide galaxies into bins of velocity dispersion log\sigv\ of width $\rm \Delta\log\sigv=0.05$ and
bins of $r$-band absolute magnitude $M_r$ of width ${\rm \Delta} M_r=0.5$. In each of such bins, galaxies
are then ordered with increasing \dn\ strength, and their spectra are stacked until a minimum S/N of 40 is
reached. Each spectrum is weighted by 1/\vmax, where \vmax\ is the maximum visibility volume given by the
bright and faint magnitude limits of the sample ($14.5\leq r\leq17.77$), and by our requirement that the galaxy redshift be included
between 0.005 and 0.22. The true number density of galaxies in the Universe should be estimated by accounting
for galaxies that are missed due to, e.g., fibre collisions and spectroscopic failures. To correct for this,
we have compared the $r$-band luminosity function obtained with our \vmax\ estimates with the luminosity
function of \cite{blanton_lf03} and derived a normalisation factor for our \vmax\ estimates. At the end we
obtain 14,694 coadded spectra from 122,643 spectra of low-S/N galaxies in the redshift interval
$0.005<z\leq0.22$.

\begin{figure}
\centerline{\includegraphics[width=9truecm]{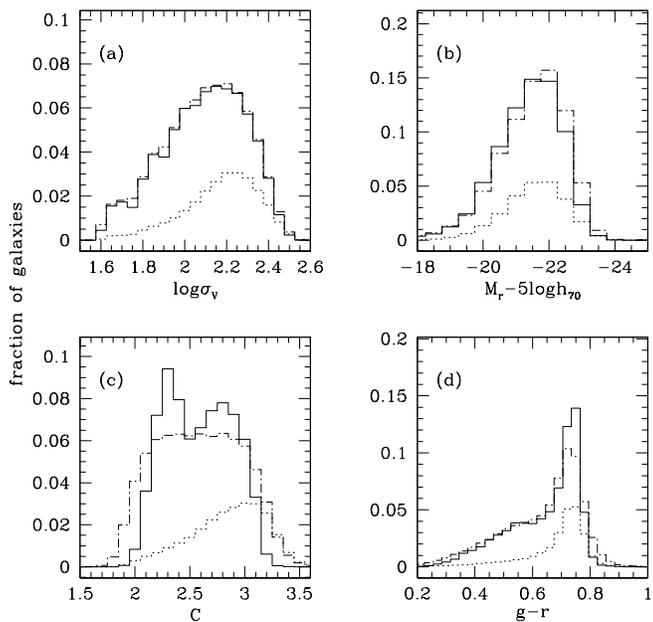}}
\caption{Distribution in velocity dispersion (a), $r$-band absolute magnitude (b), concentration
parameter (c), and ({\it k}-corrected) $g-r$ colour (d). The solid
line in each panel shows the distribution for the sample of 14,694 stacked spectra. This can be
compared to the distribution of the low-S/N galaxies, shown by the dot-dashed line. The dotted line
in each panel represents instead the distribution of the high-S/N galaxies.}\label{distr_obs}
\end{figure}

Fig.~\ref{distr_obs}a,b show the distribution in velocity dispersion and $r$-band absolute magnitude
for the coadded spectra (solid line), compared to the distribution for the individual low-S/N galaxies (dot-dashed 
line). For each stacked spectrum we estimate the absolute magnitude $M_j$ in a band $j$ as the weighted
sum of the luminosities $L_{i,j}$ of the low-S/N galaxies contributing to the coadded spectrum, according to:
\begin{equation}
{M_j = -2.5~ \log \left(\frac{\sum_i \left( L_{i,j}~w_i \right)}{\sum_i w_i} \right)+const.}\label{mag}
\end{equation}
where $w_i$ is the weight 1/\vmax\ of the individual galaxies.
The distribution in these two quantities as obtained from the stacked spectra agrees very well with the
original distribution for the low-S/N galaxies, as expected since galaxies have been binned in velocity
dispersion and absolute magnitude. It is interesting to look how well the distribution in other
morphological and photometric properties, into which galaxies are not explicitely binned, is reproduced.
Fig.~\ref{distr_obs}c,d show the distribution in the concentration parameter $C=R_{90}/R_{50}$, where
$R_{90}$ and $R_{50}$ are the $r$-band Petrosian radii, and in rest-frame $g-r$ colour. The colour of
each stacked spectrum is estimated as the difference between magnitudes defined according to
equation~\ref{mag}. The concentration parameter assigned to each stacked spectrum is given by the
1/\vmax\ weighted-average concentration parameter of the galaxies that contribute to the stacked
spectrum\footnote{Very similar results are obtained if we assign to each stacked spectrum a concentration
parameter given by the ratio between the weighted-average Petrosian radii.}. The distributions for the
stacked spectra and the low-S/N galaxies agree reasonably well, due to the correlation between colour and
velocity dispersion or magnitude, and the small scatter in concentration parameter at given log\sigv,
$M_r$ and \dn\ (the mean absolute deviation in each such bin is typically 0.18). The dotted line in each
panel of Fig.~\ref{distr_obs} shows for comparison the distribution for the high-S/N galaxies. This
clearly shows that by excluding low-S/N galaxies we would miss a substantial fraction of small,
low-concentration, blue galaxies, i.e. preferentially young, metal-poor, star forming galaxies.

We note that the distribution in concentration parameter obtained from the coadded spectra is clearly
bimodal and narrower than the distribution of the original low-S/N sample. A bimodality in $C$ is
expected from the bivariate distribution in the plane described by $C$ versus \dn. The choice of stacking
spectra with respect to \dn and the definition of weighted-average concentration parameter for the
coadded spectra give higher-S/N measures of the concentration index for `blue' sequence and `red'
sequence galaxies separately, thus enhancing the bimodality in $C$.

\begin{figure*}
\centerline{\includegraphics[width=12truecm]{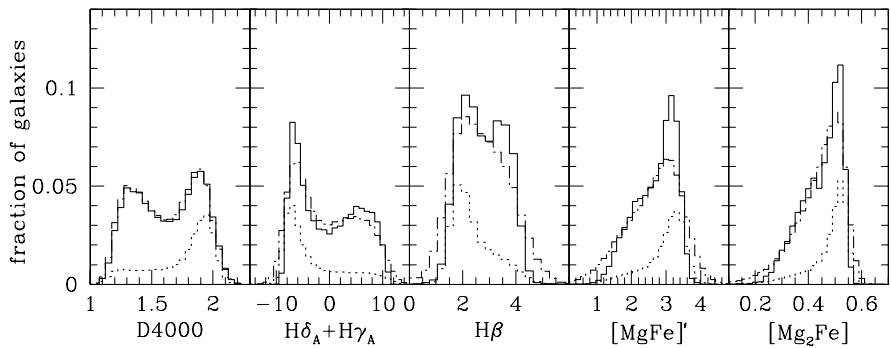}}
\caption{Distribution in the five spectral absorption features adopted to derive estimates of stellar
metallicity, age and stellar mass for the sample of 14,694 high-S/N stacked spectra (solid line),
corresponding to 122,643 low-S/N galaxies (dot-dashed line). The absorption indices are measured off the
stacked spectra in the same way as they are measured off the spectrum of individual galaxies and they
represent the 1/\vmax-weighted average index of all the galaxies that enter each stacked spectrum. Dotted
lines represent the distribution for the 42,103 high-S/N galaxies.}\label{distr_idx}
\end{figure*}

From each stacked spectrum, we also measure \dn, the higher-order Balmer lines and the other spectral
absorption indices defined in the Lick system, in the same way as they are measured from the spectrum of
individual galaxies (see also section 2.2 of paper~I). They represent the 1/\vmax-weighted average of
the absorption indices of the galaxies that contribute to each coadded spectrum. More properly,
the fluxes in the `pseudo-continuum' and central bandpasses measured from the coadded spectrum are the
1/\vmax-weighted average of the fluxes measured from the individual galaxy spectra. In
Fig.~\ref{distr_idx} the distribution in the five spectral absorption features used to constrain stellar
metallicity, age and stellar mass estimates as measured from the stacked spectra (solid line) is
compared to the distribution for the original sample of 122,677 low-S/N galaxies (dot-dashed line). The
distributions for the stacked spectra are in very good agreement with the distributions for the
original low-S/N galaxies. This is particularly true for \dn, as expected, since the spectral coaddition
is performed on galaxies with similar \dn. The comparison for the other indices shows that the increased
signal-to-noise ratio in the stacked spectra removes the tails of outliers present in the distributions
for the original low-S/N galaxies, but absent in the distributions for the 42,103 high-S/N galaxies. 

\subsection{Physical parameters estimates}\label{estimates}
Estimates of stellar metallicity, (light- and mass-weighted) age and stellar mass are derived from the
coadded spectra in the same way as they are derived from individual galaxy spectra, as summarised in
Section~\ref{sample} and more extensively described in paper~I. The physical parameters are derived by
fitting the galaxy spectra as observed and so they refer to the galaxies at the time of observation. This
concerns in particular the stellar age. In Section~\ref{distrz} and~\ref{age} we will study the
distribution of metals as a function of stellar age. In order to define a characteristic age and
interpret it as a characteristic redshift of metal production, we correct the measured mass- and
light-weighted ages by adding the lookback time to the redshift at which the galaxy is observed. The age
obtained in this way represents the effective (mass- or light-weighted) epoch when stars formed. For the
stacked spectra we assume the average redshift of all the galaxies that contribute to each spectrum. The
spread in redshift of the galaxies contributing to each coadded spectrum is on average 30 percent for
redshift up to 0.1 and 20 percent for $z>0.1$. 

\begin{figure}
\centerline{\includegraphics[width=10truecm]{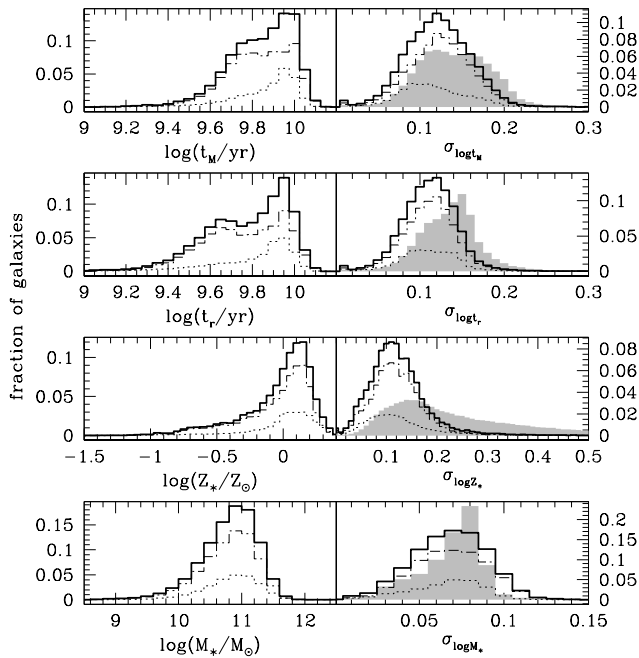}}
\caption{Distribution in mass-weighted age, $r$-band light-weighted age, stellar metallicity and stellar
mass (from top to bottom, left panels) for the final sample obtained by combining high-S/N galaxies and
the low-S/N galaxies included in the coadded spectra (thick solid line). The dotted lines show the
contribution by high-S/N galaxies only, while the dot-dashed lines represent the distribution in the
parameters for the low-S/N galaxies, as derived from the stacked spectra. The right-hand panels show the
distribution in the corresponding uncertainties, given by half of the $16-84$ percent percentile range of
the likelihood distribution. The grey-shaded histograms in the right-hand panels give for comparison the
distribution in the 68 percent confidence interval of the physical parameters of low-S/N galaxies as
derived from their individual spectrum.}\label{distr_param}
\end{figure}

The distribution in the derived parameters for the whole sample of 164,746 galaxies is shown in the
left-hand panels of Fig.~\ref{distr_param} (thick solid line). The mass-weighted age of the sample peaks
at $\sim$10~Gyr and then extends towards ages as young as 2.5~Gyr. The $r$-band light-weighted age shows
a roughly bimodal distribution with a primary peak around 9~Gyr and a broader peak around 4~Gyr. The
distribution in stellar metallicity is also highly skewed with a primary peak around $1.4\times Z_\odot$
and an extended tail towards lower metallicities. The effect is much weaker for stellar mass, which has a
distribution roughly symmetric around a mean $\log(M_\ast/M_\odot)=10.81$ with a scatter of 0.46~dex.  We
note that, as expected, the derived mass-weighted age of the galaxies is older than their light-weighted
age. This difference is larger for younger galaxies, going from 0.7~Gyr for galaxies with t$_r$=6.3~Gyr
up to 4~Gyr for galaxies with t$_r$=2.5~Gyr. This reflects (on average) the more extended star formation
histories of younger (less massive) galaxies.

The right-hand panels show the distribution in the uncertainties on mass-weighted age, $r$-band
light-weighted age, stellar metallicity and stellar mass, given by half of the $16-84$ percent percentile
range of the corresponding likelihood distribution. The dotted line shows the distribution for the
high-S/N galaxies, while the dot-dashed line shows the distribution for the uncertainties on the
parameters of low-S/N galaxies as derived from the coadded spectra. This can be compared to the 68
percent confidence range on the parameters of low-S/N galaxies as derived from their individual spectra
(grey-shaded histogram). This makes clear the importance of a good spectral S/N in the determination of
the physical parameters (in particular stellar metallicity, see also paper~I) and the advantage of the
stacking technique: it allows us to retrieve the physical parameters of galaxies with low-S/N spectra
with a much better accuracy than what we could do from their individual spectra.

\begin{figure}
\centerline{
\includegraphics[width=4truecm]{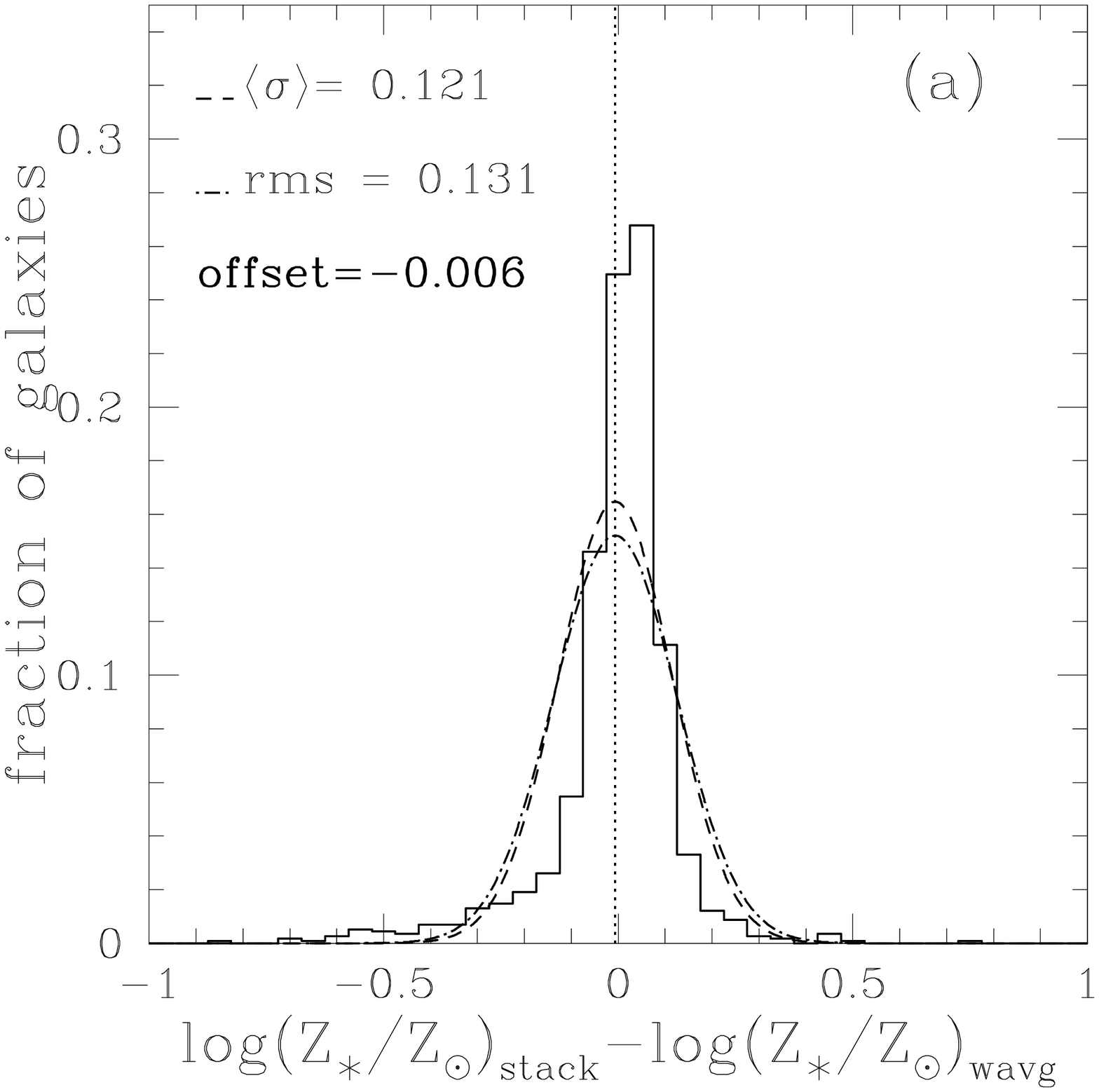}
\includegraphics[width=4truecm]{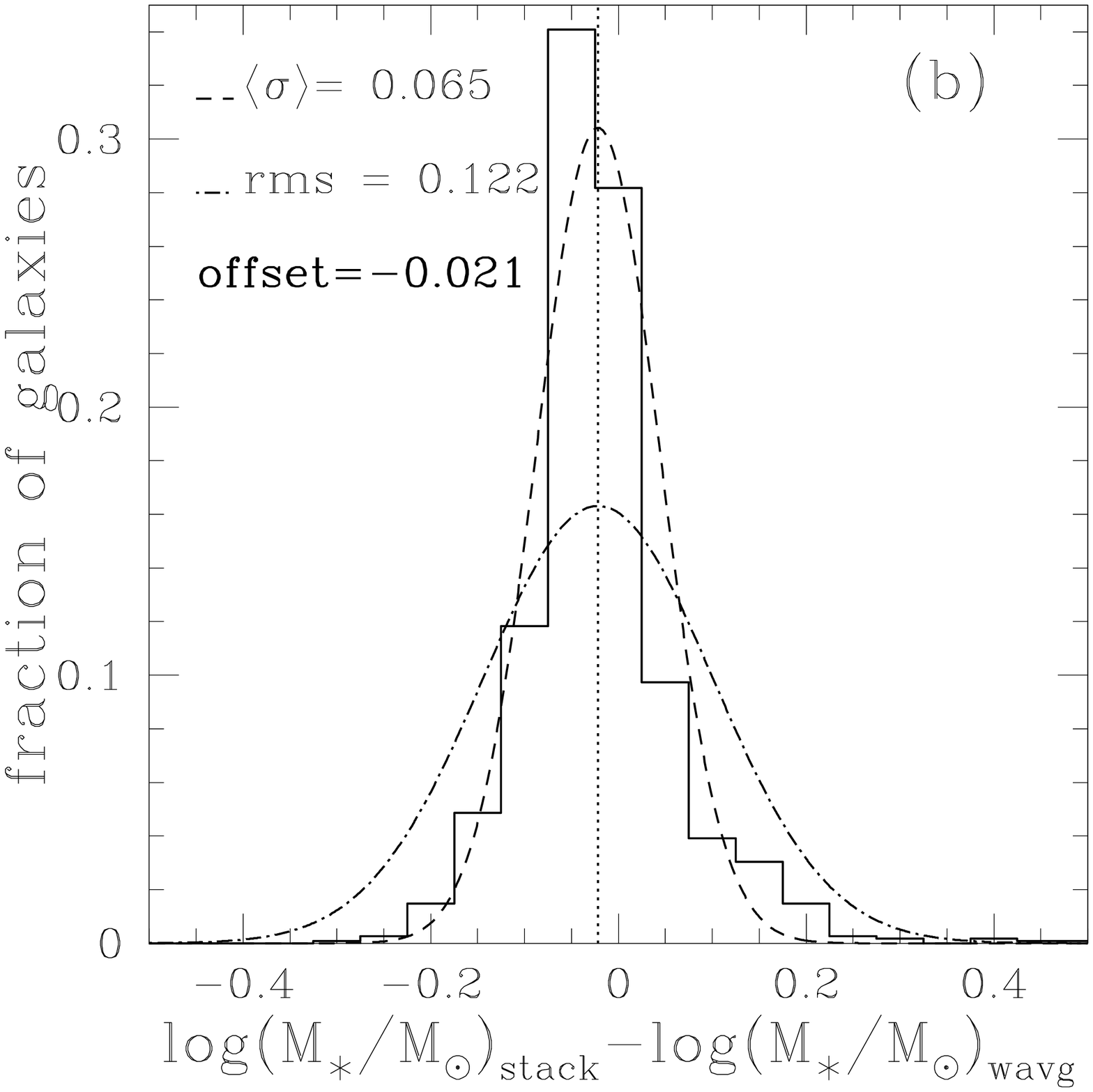}}
\centerline{
\includegraphics[width=4truecm]{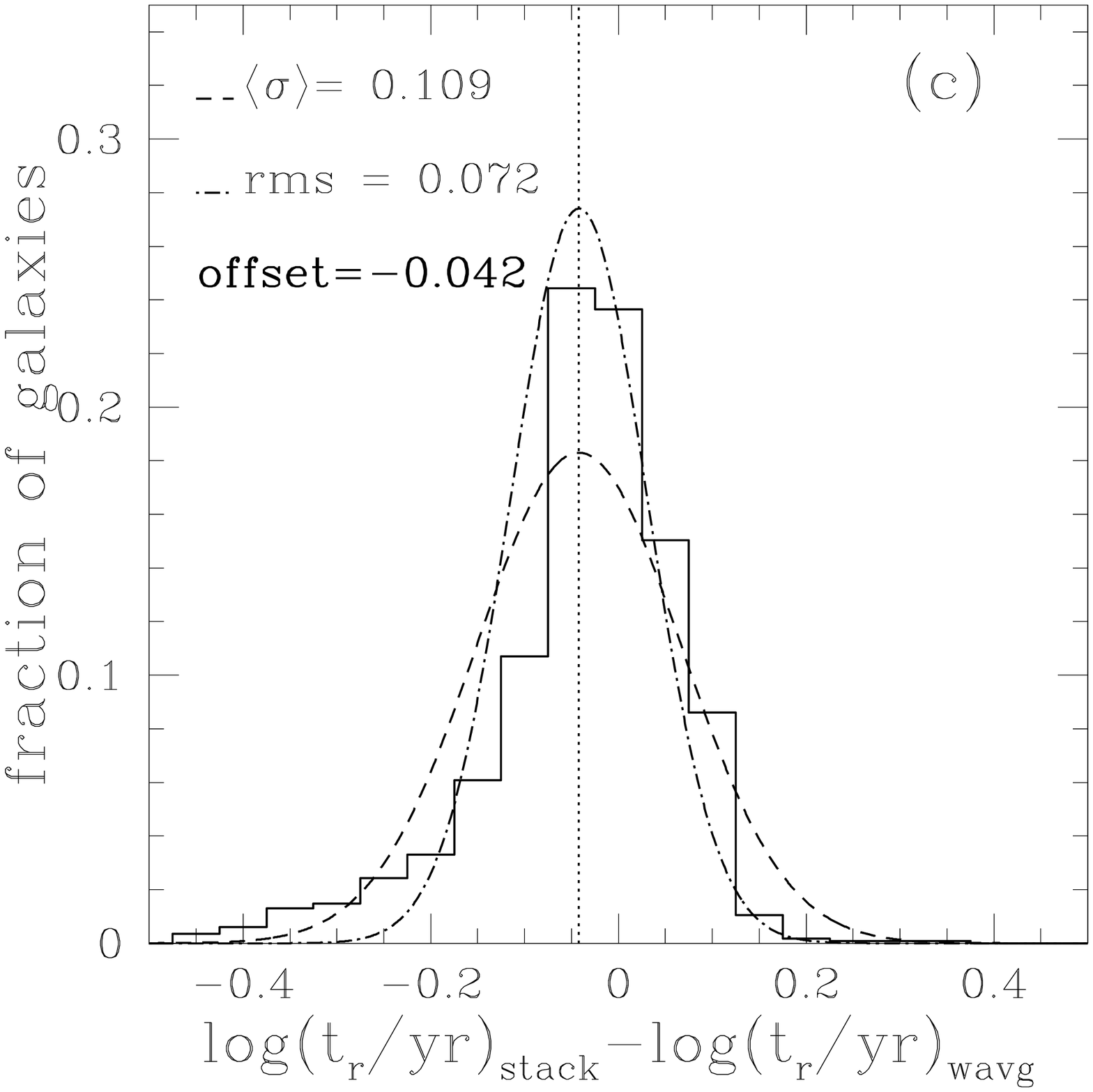}
\includegraphics[width=4truecm]{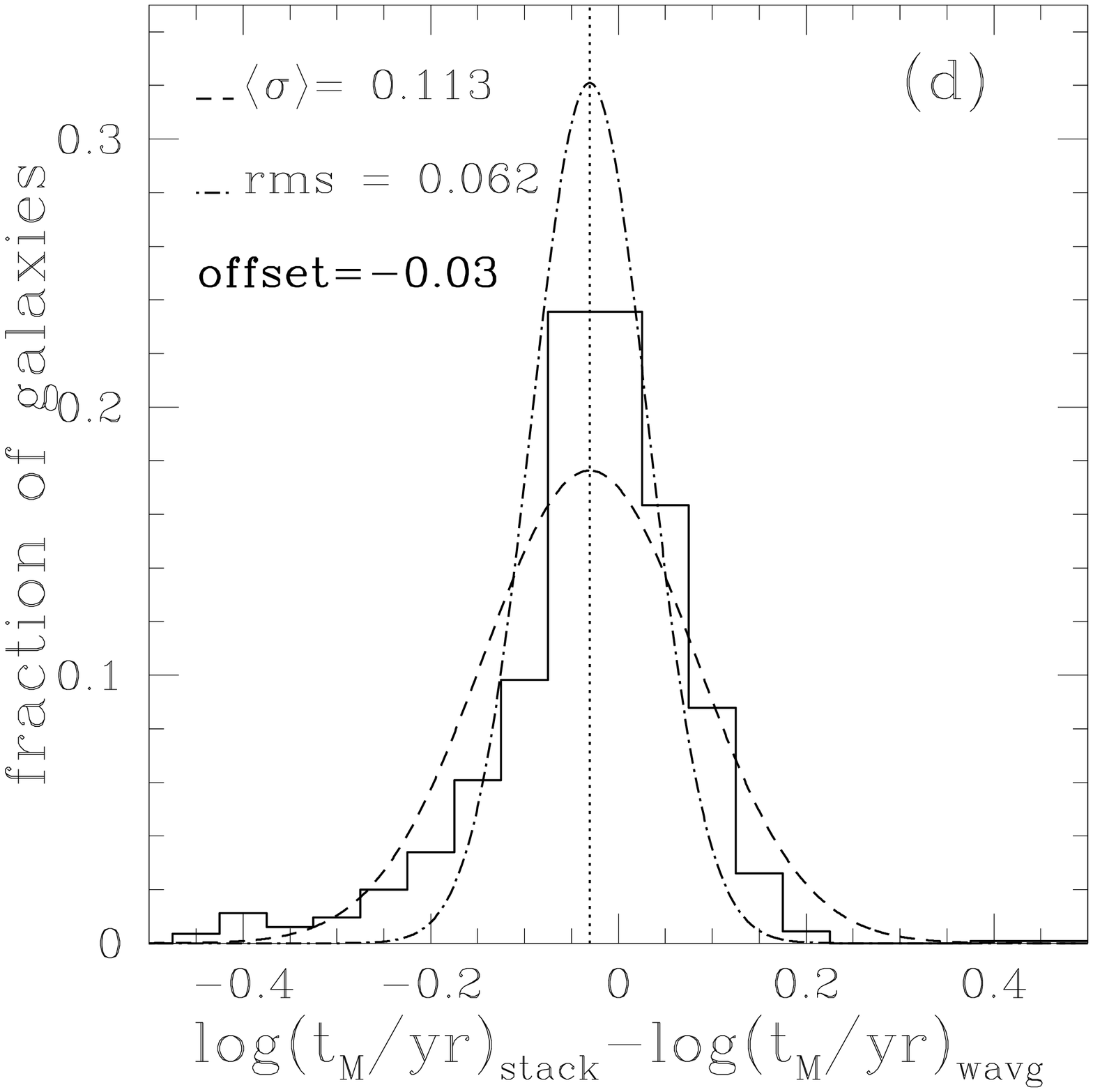}}
\caption{Distribution of the difference between the stellar metallicity (a), stellar mass (b),
light-weighted age (c) and mass-weighted age (d) estimated from the stacked spectra and the
(1/\vmax)-weighted average parameter (indicated with `wavg') of the galaxies that contribute to each
stacked spectrum. Each histogram is compared to gaussian distributions of width given by the average
uncertainty on the corresponding parameter ($\langle\sigma \rangle$, dashed line) and by the average rms
scatter in the physical parameters of the galaxies that enter each stacked spectrum (rms, dot-dashed
line). For this test we used a control sample of stacked spectra obtained by coadding the spectra of
individual high-S/N galaxies, for which reliable estimates of metallicity, mass and age can be
obtained, after degrading their quality adding gaussian noise.}\label{comp_param}
\end{figure}

The physical parameters derived from the stacked spectra can be interpreted as the (1/\vmax)-weighted
average stellar metallicity, age and stellar mass of the galaxies that contribute to each coadded
spectrum. To test how well we can recover the physical parameters of individual galaxies with our
stacking technique, we have generated a control sample of stacked spectra by coadding the spectra of
individual high-S/N galaxies, for which reliable estimates of metallicity, age and mass can be derived,
in the same way as described in Section~\ref{stacking} for the low-S/N galaxies. Before coaddition we
added gaussian noise to the individual high-S/N spectra to mimic the situation we have when coadding
low-S/N spectra. \footnote{At low S/N there is likely to be non-gaussian noise sources such as sky
subtraction problems, but our approach here should capture most of the trends.} We then compare the
physical parameters estimated from the coadded spectra with the (1/\vmax)-weighted average parameters of
the galaxies that contribute to each coadded spectrum. This is shown in Fig.~\ref{comp_param} for stellar
metallicity (panel a), stellar mass (panel b), light-weighted age (panel c) and mass-weighted age (panel
d). The histogram of the difference between the derived (`stack') and expected (`wavg') parameters is
compared to gaussian distributions of width given by the average uncertainty on the derived parameter
(dashed line) and by the average scatter in the physical parameter of the galaxies that contribute to
each stacked spectrum (dot-dashed line). For all the parameters we can recover the expected value within
the corresponding typical error. We note however that there is a small but systematic offset (dotted
vertical line) on average of about $-0.02$~dex in stellar mass, $-0.03$~dex in mass-weighted age and
$-0.04$~dex in light-weighted age (while the average offset in stellar metallicity is negligible).  This
offset likely originates from the fact that each coadded spectrum tend to be dominated by the youngest,
brightest stellar populations of the galaxies that compose it, and therefore the derived light-weighted
ages and mass-to-light ratios tend to be biased low. We take this into account as a systematic
uncertainty, as discussed in Section~\ref{errors} below. \\

\section{The mass density of baryons and metals in the local Universe}\label{metal_density}
In this section we derive an estimate of the total amount of baryons and metals locked up in stars and
the average stellar metallicity in the local Universe, by combining the contribution of individual
high-S/N galaxies and the low-S/N galaxies included in the coadded high-S/N spectra
(Section~\ref{totalz}). We also discuss and quantify systematic uncertainties in Section~\ref{errors},
and compare with various observational estimates and model predictions in the literature in
Section~\ref{literature}.

\subsection{The total stellar metallicity in the local Universe}\label{totalz}
We compute the mass density of metals in stars, $\rho_Z$, and the stellar mass
density, $\rho_\ast$, at a mean redshift of $z=0.1$, as follows:
\begin{equation}
{\rho_Z = \sum_i \left (Z_{\ast,i}~M_{\ast,i}~w_i \right ) + \sum_i \left
(Z_{\ast,i}^{st}~M_{\ast,i}^{st}~W_i^{st} \right )}\label{eqn2}
\end{equation}
\begin{equation}
{\rho_\ast = \sum_i \left (M_{\ast,i}~w_i \right ) + \sum_i \left
(M_{\ast,i}^{st}~W_i^{st} \right )}\label{eqn3}
\end{equation}
where $Z_{\ast}$ and $M_{\ast}$ are the median-likelihood estimates of the stellar metallicity and
stellar  mass of the high-S/N galaxies, and $w$ are the weights 1/\vmax. The symbols $Z_{\ast}^{st}$ and
$M_{\ast}^{st}$ refer to the stellar metallicity and mass estimated from the coadded spectra. These have
to be weighted by $W^{st}$, which is the sum of all the weights 1/\vmax\ of the low-S/N galaxies
contributing to each stacked spectrum. 

Combining the contribution of individual high-S/N galaxies and coadded spectra, we derive: 
\begin{equation}
{\rho_Z = 7.099\pm0.019^{+2.184}_{-1.943}\times10^6~h_{70}~\rm M_\odot~Mpc^{-3}}\label{eqn4}
\end{equation}
\begin{equation}
{\rho_\ast = 3.413\pm0.005^{+0.569}_{-0.554}\times10^8~h_{70}~\rm M_\odot~Mpc^{-3}}\label{eqn5}
\end{equation}
We note that the derived stellar mass density corresponds to 0.0025 times the critical density (see
Table~\ref{densities}). We thus find a stellar baryon fraction of only 7 percent, assuming a value of
0.004 for the cosmic baryon fraction.

The systematic uncertainties on these quantities (and on the average stellar metallicity $\langle
Z_\ast/Z_\odot\rangle$ derived below) are discussed in detail in Section~\ref{errors} and summarized in
Table~\ref{sys_err_2}. We calculate `corrected' mass densities using stellar mass and stellar metallicity
estimates corrected for systematics as described in Section~\ref{errors}. The uncertainties on
$\rho_\ast$ and $\rho_Z$ are then expressed as the difference between the `corrected' and the default
values. The sources of systematic uncertainties that we consider include the possible bias introduced by
the adopted stacking technique, the non-modeled dependence of index strengths on element abundance
ratios, the prior distribution of the models in the Monte Carlo library, the aperture effects and the
magnitude used to normalize the stellar mass-to-light ratio.

The systematic uncertainties quoted in equations~\ref{eqn4} and~\ref{eqn5} are the largest among those
summarized in Table~\ref{sys_err_2}. These are the aperture bias, which can lead to an overestimate of
$\sim$16 percent and of $\sim$27 percent in $\rho_\ast$ and $\rho_Z$ respectively, and the stacking
technique, which may introduce an underestimate in $\rho_Z$ of about 30 percent and less than 20 percent
in $\rho_\ast$. The use of Petrosian magnitudes for determining stellar masses would provide estimates of
both $\rho_\ast$ and $\rho_Z$ lower by less than 13 percent of those derived with model magnitudes. A
similar effect would be caused by the choice of a burst-enhanced prior. For completeness in
equations~\ref{eqn4} and~\ref{eqn5} we also indicate the statistical error (estimated by standard error
propagation from the uncertainties on metallicity and mass). This is very small and represents only the
0.2 percent of the value of $\rho_Z$ and 0.1 percent of the value of $\rho_\ast$ that we derive. 

We can now combine Equation~\ref{eqn2} and~\ref{eqn3} to estimate the (mass-weighted) average metallicity
in stars in the nearby Universe, obtaining:
\begin{eqnarray}
\langle Z_\ast\rangle & = & \frac{\sum_i \left (Z_{i}~M_{\ast,i}~w_i \right ) + \sum_i \left
(Z_i^{st}~M_{\ast,i}^{st}~W_i^{st} \right )}{\sum_i \left (M_{\ast,i}~w_i \right ) + \sum_i \left
(M_{\ast,i}^{st}~W_i^{st} \right )} \nonumber \\
& = & 1.04\pm0.002^{+0.145}_{-0.14} Z_\odot \label{eqn6} 
\end{eqnarray}
The normalization of the total metallicity in stars is affected mainly by the choice of prior and by
aperture effects, which can lead to an overestimate and underestimate, respectively, of about 14 percent.
The uncertainties introduced by the stacking technique amount to roughly 12 percent (see
Table~\ref{sys_err_2}). The scaled-solar abundance ratio of the models and the choice of normalization
magnitude can introduce systematics lower than 10 percent. We note that, when considered separately,
high-S/N galaxies give a total metallicity of $1.149 Z_\odot$ while low-S/N galaxies give, as expected, a
lower metallicity of $0.996 Z_\odot$.\footnote{Low-S/N galaxies are preferentially galaxies with low
surface brightness and low concentration parameters. As shown in paper~I, while (the bulges of) massive
disc-dominated galaxies have metallicities comparable to those of bulge-dominated galaxies, low-mass
$C\leq2.4$ galaxies are on average more metal-poor than their $C\geq2.8$ counterparts. We thus expect
low-S/N galaxies to have a mass-weighted mean metallicity slightly lower than high-S/N (mostly
bulge-dominated) galaxies. } 

We derive thus that the average metallicity of stars in the present-day Universe is the typical
metallicity of $\rm L^{\ast}$ galaxies, i.e. consistent with the solar value. We mention in passing
that, while the solar metallicity scale has been substantially adjusted downwards with respect to the
previously recommended value \citep{asplund05a,asplund05b}, this has no impact on our analysis. This is
because the BC03 models are tied to the iron abundance [Fe/H], whose solar value is unchanged with the
new calibration. In view of this we adopt $Z_\odot=0.02$ in this paper for consistency with our previous
work. 

Our result, based on a large homogeneous sample of galaxies spanning large ranges in physical
properties, and on recent population synthesis models accounting for the full range of possible SFHs,
puts on a more robust basis (and with accurate estimates of the systematic uncertainties) an early
calculation by \cite{edmunds97}, based on closed-box chemical models for irregulars, ellipticals and
spirals. Assuming mean metal abundances and combining
contribution by different galaxy types according to their different luminosity functions and average
stellar mass-lo-light ratios, they derive a mass-weighted mean oxygen abundance (including stars and gas)
of $\rm 12+\log(O/H)=8.8$, i.e. consistent with solar\footnote{This value would correspond to 1.28 times
the solar value, adopting a solar oxygen abundance of $\rm 12+\log(O/H)_\odot=8.69$
according to \cite{allende01}.}. This is also in agreement with \cite{cm2004}, who, summing the
contribution of ellipticals, spirals and irregular galaxies from a chemo-photometric model, predict a
mass-weighted mean metallicity in stars of 0.019, i.e. roughly $0.95~Z_\odot$ (from their tables
1,~2,~3,~9).

We stress that the metal mass density that we estimate here accounts only for the fraction of metals in
the stellar populations of present-day galaxies, and not those in the interstellar medium (nor in other
gaseous form outside galaxies). In the local Universe, the stellar components in galaxies host a
significant fraction ($30-50$ percent) of the cosmic metal content, almost comparable to the fraction of
metals that reside outside galaxies (according to Calura\&Matteucci 2004, and as recently confirmed, within
uncertainties, by Dav{\`e}\&Oppenheimer 2007 using cosmological hydrodynamical simulations
which incorporates enriched galactic outflows in a hierarchical structure formation scenario). 
This is not
the case at higher redshift (above $z\sim2$), when the diffuse inter-galactic gas is the largest
reservoir of metals. These considerations find support in the observational census compiled by
\cite{dunne03}.

Following naively the discussion of \cite{edmunds97} the solar average stellar metallicity in the local
Universe would be consistent with the idea that most of the baryons available for star formation have
been locked up into stars (if the effective yield has approximately solar value).\footnote{In a
closed-box system the average stellar metallicity would be approaching the yield as the gas is
progressively processed into stars.} Considering only galactic components, while at high redshift the
majority of baryons are in the ISM, at low redshift the largest reservoir of baryons are the stars
\citep[see e.g.][]{dunne03,cm2004}. This is consistent with the decline in the ensemble star formation
rate since $z\sim 1$ till the present. However, the majority of {\it all} baryons appears to be at all
epochs outside galaxies, in the diffuse IGM \citep[e.g.][]{DO07}.

\subsection{Sources of uncertainties}\label{errors}
We discuss here the possible sources of systematic uncertainties that we identify in our procedure for
deriving physical parameters estimates from individual or coadded galaxy spectra (Tables~\ref{sys_err_1a}
and~\ref{sys_err_1b}), and quantify how they propagate into the estimates of stellar mass density, mass
density of metals in stars, and average metallicity (Table~\ref{sys_err_2}). The results quoted in
Equations~\ref{eqn4},~\ref{eqn5},~\ref{eqn6} are {\it not} corrected for systematics. For each source of uncertainty,
we estimate here, as best as we can, corrections on individual stellar masses and metallicities (and
ages), and calculate `corrected' $\rho_\ast$, $\rho_Z$ and $\langle Z_\ast\rangle$. The systematic
uncertainties on the these quantities are then expressed as the difference between their value after
correction and the default one (as given in Section~\ref{totalz}). Fig.~\ref{t50_mass} in
Section~\ref{age} provides a visual representation of the extent of each systematic uncertainty as a
function of galaxy stellar mass.
\\

{\it Stacking technique.}
We already mentioned that the stacking technique we adopt may provide physical parameters estimates which
are (slightly) systematically lower (on average) than the expected average parameters of low-S/N
galaxies. This affects in particular stellar mass and mean ages. To quantify the extent of these effects
we measured the range between the upper and lower quartiles of the distributions shown in
Fig.~\ref{comp_param}, i.e. the difference between measured and expected parameters (first row of
Table~\ref{sys_err_1a}). We do this also for stellar metallicity although the average offset is
negligible. As mentioned before, while the differences in stellar metallicity are symmetric around zero
and well within the typical uncertainty on metallicity, the masses and ages estimated from the coadded
spectra are in general lower than the expected parameters by an amount corresponding to $\sim$30 percent
of the typical uncertainties on these parameters. We correct stellar metallicity and mass estimates
according to these offsets and calculate `corrected' $\rho_\ast$, $\rho_Z$ and $\langle Z_\ast\rangle$.
The systematic uncertainties on these quantities are then expressed as the (interquartile range of the)
difference between the `corrected' and default values (Table~\ref{sys_err_2}).
\\

{\it $\alpha$/Fe abundance ratio.}
The Monte Carlo library we use is based on the BC03 population synthesis code, which does not
include the effect of variations in element abundance ratios with respect to solar. In deriving galaxy
physical parameters we take care to constrain only those absorption indices which have a weak
dependence on $\alpha$/Fe abundance ratio. Nevertheless, recent studies have shown that the higher-order
Balmer lines (also used in our analysis) are sensitive to $\alpha$-enhancement at metallicities around
solar and above and may lead to an underestimate of stellar ages if models with scaled-solar abundance ratios
are used \citep{thomas04,korn05,prochaska07}. We estimate the possible (mass-dependent) systematic
uncertainty on the physical parameters we derive with scaled-solar abundance ratios models as follows. 

First of all, for a stellar population with abundance ratio $\rm [\alpha/Fe]=0.3$, we determine the
offset  between the expected (true) parameters and those derived with a model library with $\rm
[\alpha/Fe]=0$, using the predictions from \cite{thomas04} simple stellar populations (SSP) models. This
is described in detail in section 2.4.2 of paper~I. The second row of Table~\ref{sys_err_1a} gives the
interquartile range of the offsets (for $\rm [\alpha/Fe]=0.3$) in stellar metallicity, light- and
mass-weighted age, and stellar mass-to-light ratio.

The degree of $\alpha$-enhancement in ellipticals is a function of the galaxy stellar mass. Therefore we
derive mass-dependent corrections to the physical parameters on a galaxy-by-galaxy basis and quantify the
effect on $\rho_\ast$, $\rho_Z$ and $\langle Z_\ast\rangle$. We cannot derive an estimate of the true
$\alpha$/Fe abundance ratio of individual galaxies with our models. We use instead an empirical estimate
given by $\rm \Delta(Mgb/\langle Fe\rangle)$, i.e. the difference between the observed $\rm Mgb/\langle
Fe\rangle$ index ratio and the index ratio of the $\rm [\alpha/Fe]=0$ best-fit model in our library (see
section 2 of paper~II). We then use the relationship between $\rm \Delta(Mgb/\langle Fe\rangle)$ and
stellar mass derived for bulge-dominated galaxies ($C\geq2.8$, see table 4 of paper~II). This gives us
the degree of $\alpha$/Fe expressed in terms of the galaxy stellar mass. From this we estimate the
correction to be applied to the galaxy physical parameters, scaling the offsets derived above
(Table~\ref{sys_err_1a}) accordingly. We apply these corrections only to $C\geq2.8$ galaxies. 
\\

{\it Prior.}
Another possible source of systematic error comes from the choice of prior according to which
our model library populates the parameter space, in particular the mixture of bursty and continuous star
formation histories. As discussed in section 2.4.2 of paper~I, we explored the effect on the derived
physical parameters of changing the fraction of bursts in the last 2~Gyr from 10 percent (our standard
prior) to 50 percent. The mean ages and the stellar mass derived with this burst-enhanced prior are on
average lower than those derived with our standard prior, while the stellar metallicity is slightly
higher. We quantify this as the interquartile range of the difference between our standard prior and the
burst-enhanced one in stellar metallicity, light- and mass-weighted age, and stellar mass (third row of
Table~\ref{sys_err_1a}). As above, we obtain the range of the corresponding uncertainty on $\rho_\ast$, $\rho_Z$ and $\langle
Z_\ast\rangle$ as the difference between the `corrected' and default values.
\\

{\it Aperture effects.}
Estimates of stellar metallicity and mean stellar age are affected by the aperture bias, due to the fact
that the SDSS spectra sample only a limited inner region of the galaxy. The light collected by the fibre
is on average 30 percent of the total flux, but this fraction depends on the stellar mass, morphology and
redshift of the galaxy. Due to the presence of stellar populations gradients, the metallicities and ages 
derived from the SDSS spectra are not representative of the galaxy as a whole but only of the bulge or
central regions. A correction for this would require an accurate knowledge of metallicity gradients as a
function of galaxy type and mass, and is not feasible here. This is clearly a concern in this work, where
we want to estimate the {\it total} amount of metals in stars today (see also section 3.4 of paper~I for
a discussion on aperture effects).

We attempt to quantify this bias for galaxies with similar stellar mass and concentration index by
looking at trends in their estimated physical parameters as a function of the fraction of light in the
fibre (given by the ratio $f$ between the fibre and the Petrosian flux). Stellar mass estimates can also
be affected by aperture bias, because they are based on the (potentially wrong) assumption that the
stellar mass-to-light ratio outside the fibre is the same as the one inside (derived from the galaxy
spectrum). In the same stellar mass and concentration bins as above we checked and quantified also the
trend in the $z$-band stellar mass-to-light ratio $\log (M_\ast/L_z)$ with $f$.\footnote{For reference, a
flux ratio of about 50 percent occurs when the fibre radius coincides with the effective Petrosian radius
$\rm R_{50}$ for a de~Vaucouleur profile or is $\rm 1.3\times R_{50}$ for an exponential profile. When
the fibre covers roughly twice $\rm R_{50}$, the flux gathered is about 80 percent and 60 percent for a
de~Vaucouleur and exponential profile respectively.}  The linear relations fitted are summarized in
Table~\ref{sys_err_1b} for those stellar mass and concentration bins where the trends are statistically
significant. The relatively large variation in metallicity that we find for bulge-dominated galaxies
are, at least qualitatively, consistent with the typical abundance gradients found in the literature
\citep[$\sim0.16-0.2$dex/decade in radius, e.g.][]{henry99,mehlert03}. We note that we are not able to
identify significant trends in stellar metallicity for $C<2.4$ galaxies. It is possible that we are
underestimating the aperture effects for this class of galaxies. However, while abundance gradients are
observed in individual spiral galaxies, they are on average rather shallow and associated instead to
larger age gradients \citep{macarthur04,BdJ01}, consistent with our findings.

For each galaxy, we calculate new physical parameters estimates by adding the
variation in the corresponding parameter obtained by extrapolating the relations given in
Table~\ref{sys_err_1b} to $f=1$. We then recalculate the total $\rho_\ast$, $\rho_Z$ and $\langle
Z_\ast\rangle$. As above, the systematic uncertainty on these quantities is expressed as the difference
between the `corrected' and default values.
\\

{\it Total magnitude.}
The stellar mass estimates are influenced by the total magnitude we choose as
normalization of the stellar mass-to-light ratio. We use the SDSS model magnitude\footnote{This is the
total magnitude of the best-fit model between a de~Vaucouleurs and an exponential profile to the $r$-band
image of the galaxy.} as a good estimate of the total galaxy luminosity, with respect to the more often
adopted Petrosian magnitude \citep[e.g. by][]{kauf03a}, measured within a circular aperture of radius two
times the Petrosian radius. The Petrosian flux should recover almost all the flux for an exponential
profile and $\sim 80$ percent of the flux for a de~Vaucouleurs profile. For completeness and for
comparison with other works, we also derive stellar mass estimates using the $z$-band Petrosian
luminosity to normalize the $M_\ast/L_z$. The last row of Table~\ref{sys_err_1a} gives the interquartile
range of the difference between `model' and `Petrosian' stellar masses.
\\

Finally, we note that the adopted shape of the initial mass function (IMF) and the assumption that
it is universally applicable between galaxies with different mass and star formation history clearly have
an impact on the derived stellar mass, and hence $\rho_\ast$ and $\rho_Z$. Exploring in details the
effects of the different assumptions goes beyond the scope of the present work. However, we try here to
quantify how much the mass-to-light ratio scale would vary by varying the parameters of the IMF. In
particular we consider the effect of changing the Chabrier IMF parameters within their quoted
uncertainties \citep[see Table 1 of][]{chabrier03}. We calculated the evolution of the $z$-band
mass-to-light ratio for SSPs of different metallicity (from 0.4 to 2.5 times solar) for two `modified'
Chabrier IMFs, one with the steeper high-mass slope of 1.6 and upper limits for the low-mass end
parameters, and another one with the shallower high-mass slope of 1 and lower limits for the low-mass end
parameters. In the first case the ratio of $M_\ast/L_z$ between the modified and the standard Chabrier
IMF varies from 0.73 at 1~Gyr to 0.66 at 10~Gyr. In the second case this ratio varies from 1.55 to 1.75
over the same time interval. In both cases there is negligible dependence on metallicity. We also note
that the difference in $M_\ast/L_z$ between the various IMFs roughly corresponds to the difference
between SSPs of different metallicities at a given IMF. Similar results are obtained when looking at
fixed absorption index strength, such as \dn\ or \mgfep\ or \hdg, rather than at fixed age. We note that
the the uncertainty in the slope at the high-mass end is largely responsible of the  significant
variation in $M_\ast/L_z$ between the `modified' and `standard' IMFs,  by varying the weight of stars
just above the turnoff mass of $\rm 1M_\odot$. If we let vary only the low-mass end parameters, fixing
the high-mass slope at $1.3$, the $M_\ast/L_z$ is $\sim1.03$ times and $\sim0.97$ times (for `lower' and
`upper' case respectively) the $M_\ast/L_z$ of the standard Chabrier IMF.  Since it will be useful in
Section~\ref{literature}, we also mention here the comparison with the single power-law \cite{salpeter}
IMF. For a solar metallicity simple stellar population the ratio between the $z$-band $M_\ast/L$
predicted by the Salpeter IMF and the one predicted by the Chabrier IMF is on average 1.75 and varies
from 1.72 at 1~Gyr to 1.76 at 10~Gyr.

\begin{table*}
\centering
\caption{Summary of the uncertainties or parameter choices that can affect the stellar metallicity
(column 2), $r$-band light-weighted age (column 3), mass-weighted age (column 4) and stellar mass (column
5)  estimates. The different entries, indicated in the first column, are: the offset between the
parameters derived from the stacked spectra and the expected 1/\vmax-weighted average parameters of the
corresponding low-S/N spectra; sensitivity to non-solar element abundance ratios, not taken into account
in BC03 models; the prior according to which the models of the Monte Carlo library are distributed in
parameter space (in particular we consider a library in which 50 percent of the models had a burst in the
last 2~Gyr against our default library in which this fraction is only 10 percent); the total $z$-band
galaxy magnitude used to normalize the stellar mass-to-light ratio of the models (in particular we
consider the difference of using Petrosian magnitude instead of our default choice of using model
magnitude). The extent of these uncertainties is expressed as the upper and lower quartiles of the
difference between the measured value and the expected (corrected) one. }\label{sys_err_1a}
\begin{tabular}{|crrrr|}
\hline
 & $\Delta\log(Z_\ast/Z_\odot)$ & $\Delta\log(t_r/yr)$ & $\Delta\log(t_M/yr)$ & $\Delta\log(M_\ast/M_\odot)$ \\
(1) & (2) & (3) & (4) & (5) \\ 
\hline
stacking & $-0.070 : 0.036$ & $-0.108 : 0.000$ & $-0.099 : 0.011$ & $-0.091 : -0.011$  \\ 
$\alpha$/Fe & $-0.084 : 0.146$ & $-0.166 : -0.003$ & $-0.170 : 0.026$ & $-0.119 : -0.014$ \\
prior     & $-0.057 : -0.010$ & $0.059 : 0.103$ & $0.051 : 0.091$ & $0.033 : 0.067$   \\
total magnitude &$-$ &$-$ &$-$ & $0.019 : 0.071$  \\
\noalign{\smallskip}
\hline
\end{tabular}
\centering
\caption{We quantify here the potential bias due to the fixed fibre aperture. We express this as the
variation in each physical parameter as a function of the fraction of light missed by the fibre ($\Delta
f$) at given stellar mass and concentration parameter (column 1). In the table we give the slopes of
those relations which are statistically significant. (See text for more details.)}\label{sys_err_1b}
\begin{tabular}{|lrrrr|}
\hline
Range in $C$ and $\log M_\ast$ & $\Delta\log(Z_\ast/Z_\odot)$ & $\Delta\log(t_r/yr)$ & $\Delta\log(t_M/yr)$ & $\Delta\log(M_\ast/L_z)$ \\
(1) & (2) & (3) & (4) & (5) \\ 
\hline
$C\geq2.8$ \& $10<\log(M_\ast/M_\odot)\leq10.3$   & $-$ & $-$ & $-$ & $-0.32\pm0.13$ $\Delta f$ \\
$C\geq2.8$ \& $10.3<\log(M_\ast/M_\odot)\leq10.5$ & $-0.08\pm0.08$ $\Delta f$ & $-$ & $-$ & $-0.21\pm0.02$ $\Delta f$ \\
$C\geq2.8$ \& $10.5<\log(M_\ast/M_\odot)\leq10.8$ & $-0.14\pm0.06$ $\Delta f$ & $-$ & $-$ & $-0.13\pm0.02$ $\Delta f$ \\
$C\geq2.8$ \& $10.8<\log(M_\ast/M_\odot)\leq11$   & $-0.14\pm0.02$ $\Delta f$ & $-$ & $-$ & $-0.20\pm0.03$ $\Delta f$ \\
$C\geq2.8$ \& $\log(M_\ast/M_\odot)>11$           & $-0.19\pm0.03$ $\Delta f$ & $-0.14\pm0.03$ $\Delta f$  &$-0.11\pm0.02$ $\Delta f$  & $-0.16\pm0.02$ $\Delta f$  \\
$2.4<C<2.8$ \& $10<\log(M_\ast/M_\odot)\leq10.3$  & $-$ & $ 0.17\pm0.12$ $\Delta f$ & $-$ & $-$ \\
$2.4<C<2.8$ \& $10.3<\log(M_\ast/M_\odot)\leq10.5$& $-$ & $-$ & $-$ & $-0.07\pm0.03$ $\Delta f$ \\
$2.4<C<2.8$ \& $10.5<\log(M_\ast/M_\odot)\leq10.8$& $ 0.11\pm0.04$ $\Delta f$ & $-$ & $ 0.14\pm0.08$ $\Delta f$  & $-0.09\pm0.02$ $\Delta f$  \\
$2.4<C<2.8$ \& $10.8<\log(M_\ast/M_\odot)\leq11$ & $-$ & $-$ & $-$ & $-0.13\pm0.02$ $\Delta f$ \\ 
$2.4<C<2.8$ \& $\log(M_\ast/M_\odot)>11$          & $-0.12\pm0.04$ $\Delta f$ & $-$ & $-$ & $-0.21\pm0.05$ $\Delta f$ \\
$C\leq2.4$ \& $10.3<\log(M_\ast/M_\odot)\leq10.5$ & $-$ & $-$ &$ 0.23\pm0.15$ $\Delta f$ & $-$ \\
$C\leq2.4$ \& $10.5<\log(M_\ast/M_\odot)\leq10.8$ & $-$ & $-$ & $-$ & $-$ \\
$C\leq2.4$ \& $10.8<\log(M_\ast/M_\odot)\leq11$   & $-$ & $-0.39\pm0.11$ $\Delta f$ & $-0.17\pm0.05$ $\Delta f$ & $-0.12\pm0.06$ $\Delta f$ \\
$C\leq2.4$ \& $\log(M_\ast/M_\odot)>11$           & $-$ & $-$ & $-0.07\pm0.12$ $\Delta f$ & $-0.04\pm0.02$ $\Delta f$ \\	
\hline
\end{tabular}
\centering
\caption{Systematic uncertainties on the total stellar mass density ($\rho_\ast$), the mass density of
metals in stars ($\rho_Z$) and the average stellar metallicity ($\langle Z_\ast/Z_\odot\rangle$). We
first estimated these quantities using the stellar metallicities and stellar masses corrected according
to Tables~\ref{sys_err_1a} and~\ref{sys_err_1b}. The systematic uncertainties are then expressed as the difference between the
`corrected' and default values.}\label{sys_err_2}
\begin{tabular}{|crrr|}
\hline
 & $\Delta\rho_\ast (\times 10^8)$ & $\Delta\rho_Z(\times 10^6)$ & $\Delta\langle Z_\ast/Z_\odot\rangle$ \\
(1) & (2) & (3) & (4) \\ 
\hline
stacking         & $0.569 : 0.063$ & $2.184 : -0.272$ & $0.125 : -0.060$ \\ 
$\alpha$/Fe      & $0.176 : 0.020$   & $0.815 : -0.437 $  & $0.065 : -0.070$  \\
prior            & $-0.249 : -0.488$ & $0.403 : -0.873 $  & $0.145 : 0.025$  \\
aperture         & $-0.554$             & $-1.943$         & $-0.14$ \\ 
total magnitude  & $-0.419$        & $-0.922$         & $-0.010$ \\
\noalign{\smallskip}
\hline
\end{tabular}
\end{table*}

\subsection{Comparison with the literature}\label{literature}
\begin{figure}
\centerline{\includegraphics[width=9truecm]{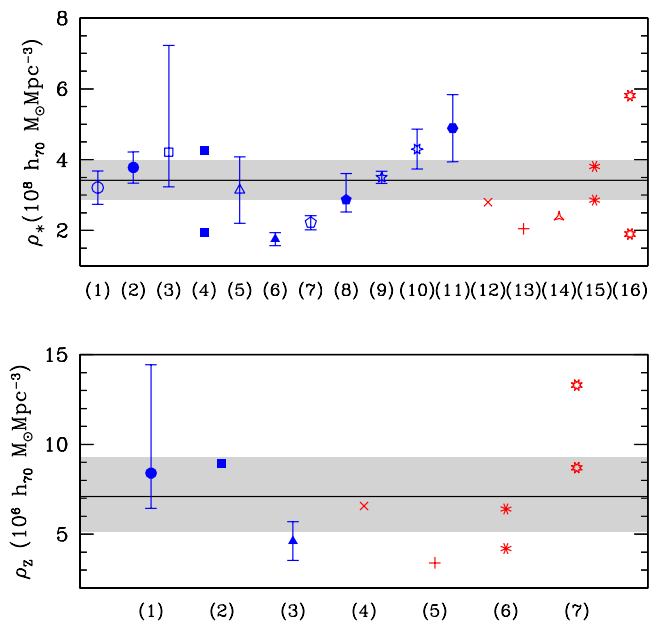}}
\caption{The estimates of metal and baryon density in stars ($\rho_Z$, $\rho_\ast$) derived in this work (solid line and
grey-shaded region) are compared to other observational determinations in the literature (blue symbols)
and to model predictions (red symbols). The number on the x-axis gives the reference according to the
caption of Tables~\ref{metal_densities} and~\ref{mass_densities}.}\label{comp_lit}
\end{figure}

We list in Tables~\ref{metal_densities} and\ref{mass_densities} the present-day mass densities of baryons
and metals in stars estimated in this work together with the values derived from several sources in the
literature, distinguishing observational estimates and prediction from models or integration of the
cosmic star formation history. All the values in the literature have been converted into our adopted
cosmology and to a Chabrier IMF, when necessary\footnote{Following \cite{bell03} we increased by 0.15~dex
the mass densities derived with a `diet'-Salpeter IMF \citep[i.e. a Salpeter IMF modified to have a flat
slope below 0.6~M$_\odot$][]{BdJ01} to convert them into a Salpeter IMF. We then assumed Salpeter-based
stellar mass densities to be 1.75 times higher than Chabrier-based densities (see last paragraph of
Section~\ref{errors}).}. The comparison is also illustrated in Figure~\ref{comp_lit}. Overall we find
good agreement within our uncertainties with other observational estimates and model predictions. 

As far as the stellar mass density is concerned, our determination is in line with the values derived by
\cite{kochanek01} from the $K$-band luminosity function of a sample of $\sim$5000 2MASS galaxies, by
\cite{cole2001} from the NIR luminosity function of a 2MASS and 2dFGRS combined sample of
infrared-selected galaxies, and similarly by \cite{bell03} from integration of the stellar mass function
derived from optical-to-NIR SED fitting on a matched sample of SDSS/2MASS galaxies. We find good
agreement also with \cite{ben06}, who apply the MOPED algorithm to fit the entire optical spectrum of
SDSS DR3 galaxies, based on BC03 models. This is encouraging given the similarity in the approach and in
sample definition. The offset of $\sim$0.07~dex is accounted for by the different definition of `total'
magnitude adopted (i.e. $z$-band model versus $r$-band Petrosian magnitude).

Analysis based on the Millennium Galaxy Catalogue also provides stellar mass density estimates in
agreement with ours, as shown in \cite{driver07a}. Recently, they revised their determination, by
incorporating effects of internal dust extinction based on an empirical relation between internal
attenuation and inclination in galaxy discs and their associated bulges \citep{driver07b}. This
correction increases by $\sim$20 percent their previously derived stellar mass density. Our estimates of
$\rho_\ast$ are still in agreement within the combined uncertainties. Similar considerations hold for the
value obtained by \cite{fhp98} combining information on the luminosity densities and stellar
mass-to-light ratios of spheroids, discs and irregular galaxies.

\cite{rudnick03}, based on a sample of SDSS EDR luminous galaxies, obtain instead a stellar mass density
roughly 50 percent lower than our value; this discrepancy is only slightly alleviated with their more
recent estimate accounting for correction to `total' values \citep{rudnick06}. Finally, our estimate of
stellar mass density falls in the range constrained by \cite{glazebrook03}, by fitting the SDSS-based
`cosmic optical spectrum' with a power-law SFH up to $z=1$.

We find marginal agreement also with the value of stellar mass density
estimated by \cite{shankar06}. Their mass-to-light ratio estimates are based on a kinematic decomposition
for late-type galaxies \citep[following][]{sp99} and on central velocity dispersion for early-type
galaxies \citep{bernardi03}, without any assumption on IMF.

Comparing our estimate of $\rho_\ast$ to model predictions, we find reasonably good agreement with the
range of values found by \cite{nagamine06}, who explore four different approaches: a two-component Fossil
model, an Eulerian hydrodynamic simulation, the analytic SFH of \cite{HS03} based on their cosmological
smoothed particle hydrodynamics simulations, and the semianalytic model of \cite{cole2000}. An early
result by \cite{pei99}, based on a set of coupled equations linking the evolution of the densities of
stars, gas, heavy elements and dust based on data from quasar absorption-line surveys, optical imaging
and redshift surveys available at that time, is also in agreement within 1$\sigma$ with our estimate. We
find instead a discrepancy of about 40 percent with the stellar mass density derived by the models of
\cite{cm2004}. A value lower by about 30 percent is presented by \cite{monaco07} based on the MORGANA
code for the formation and evolution of galaxies and AGN. Finally, we compare our measured stellar mass
density with the result of integration of the cosmic star formation history. The integration (accounting
for the fraction of mass returned to the ISM by evolved stars) of the analytic fit of \cite{cole2001} to
the SFH data of \cite{steidel99} clearly underestimate the total stellar mass density at the present
epoch. On the other hand, integration of the dust-corrected SFH provides too high a stellar mass density
(up to 70 percent higher). This discrepancy might be related to the specific dust-correction
applied. We note that \cite{bell07}, accounting for both unobscured (from UV) and dust-obscured (from IR)
star formation, find agreement between the integral of the SFH and the measured stellar mass density. In
Section~\ref{millennium} we will compare in detail our results with the predictions from the Millennium
Simulation. We anticipate here that we find good agreement between the SDSS-derived stellar mass density
and the stellar mass density predicted by the models, which amounts to $\rm 3.203\times10^8
M_\odot~Mpc^{-3}$.

There are relatively fewer determinations of the mass density of metals {\it in stars} at the present
epoch. We find agreement within 1$\sigma$ with the census by \cite{dunne03}, and with the estimate by
\cite{pagel02} based on the stellar mass density of \cite{fhp98} and assuming solar metallicity for
stars. In contrast, \cite{FP04} derive a metal mass density of only $4.6\times10^6
h_{70}~\rm M_\odot~Mpc^{-3}$ in main-sequence stars and substellar objects (their table~3), i.e. roughly 60
percent of the total metal mass density in stars that we obtain. 

Cosmic hydrodynamic simulations of \cite{DO07} predict a total metal mass density in stars of only $\rm
3.4\times10^{6} M_\odot~Mpc^{-3}$, which is at the lower end of the observational range. On the other
side, integration of dust-corrected SFH data points overpredicts the amount of metals presently locked up
into stars (as it overpredicts the stellar mass density), if we assume $\dot{\rho}_Z=y \dot{\rho}_\ast$
with $y=0.023$ \citep{madau96}. Assuming a lower yield of $y=0.015$, following a recent estimate by
\cite{conti03}, brings the integration of the cosmic chemical enrichment history in better agreement with
observations. The integration of the cosmic SFH predicted by the models of \cite{pei99} provides a better
agreement with observations, but a higher metal yield seems to be favoured in this case. A similarly
good agreement is found with the predictions from \cite{cm2004}, who study the mean metal abundances for
galaxies of different morphological types by means of chemo-photometric models for ellipticals, spirals
and irregular galaxies.\footnote{To convert their metals density from Salpeter IMF to Chabrier IMF we
adopt a factor of 1.05 (instead of 1.75 used for the stellar mass density), which accounts for the higher
metallicities expected with a Chabrier IMF. While Salpeter-based mass-to-light ratios are higher on
average by 1.75 than Chabrier-based ones, the metallicity is a factor 0.6 lower due to the fewer massive
stars ($\rm >8M_\odot$) with respect to Chabrier.} These models differ from cosmic chemical evolution
models \citep[such as those by][]{pei95,pei99} in that they are designed to study different galaxy types
and abundances of single elements, rather than average properties of the Universe. Finally, we also
calculate the metal density in stars predicted by the Millennium Simulation, obtaining a value of $\rm
4.053\times10^6 M_\odot~Mpc^{-3}$. This is $\sim1.5\sigma$ lower than our SDSS-based result (see
Section~\ref{millennium}).

As a general remark, it seems that hydrodynamical simulations have greater difficulties (with respect to
chemical evolution models) in predicting the total metal abundance, probably due to the sensitivity to
the feedback efficiency. Indeed, as far as the average stellar metallicity is concerned, while the
compilation from \cite{cm2004} provides a mean metal abundance in stars in agreement with our estimate,
recent hydrodynamical simulations seem to predict lower values of $\rm \langle Z_\ast \rangle$.
\cite{kobayashi06} propose hypernova feedback as an extra source of feedback to reach agreement between
their simulations of cosmic chemical enrichment and the observed fraction of baryons in stars (less than
10 percent; we find a stellar baryon fraction of 7 percent). However, this prescription would provide
average stellar metallicity of about $ \rm 0.7~Z_\odot$, lower than our value, even accounting for
systematic uncertainties. Similarly, simulations from \cite{DO07}, which include a self-consistent
treatment of enriched outflows, predict a mean stellar metallicity today of half solar, roughly 3$\sigma$
lower than what we derive. In their simulations, the star-forming gas in galaxies reaches instead a mean
metallicity slightly above solar by the present epoch. A higher average metallicity in the
star-forming gas with respect to stars might be expected because the gas-phase abundance traces the
abundance of the last (most metal-rich) generations of stars and not the average abundance of the entire
stellar population. Combining gas-phase oxygen abundances of SDSS galaxies from \cite{christy04},
weighting each galaxy by its gas mass\footnote{The gas masses are computed from the SFR surface density
following \cite{christy04}.} and 1/\vmax, we find an average gas-phase metallicity of
$12+\log(O/H)=8.93$. However, we cannot robustly assess that the gas-phase metallicity is indeed higher
than the stellar one, because of uncertainties in the solar oxygen abundance scale \citep{asplund05a},
and a potential systematic overestimate introduced by strong-line metallicity indicators
\citep[e.g.][]{bresolin04}.

\begin{table}
\centering
\caption{Present-day mass density of metals in stars derived from this work and from the
literature. All values have been adjusted to the concordance $\Lambda$ CDM cosmology with $H_0=70~h_{70}~\rm
km~s^{-1}~Mpc^{-1}$, with $\rho_{crit}=1.36\times10^{11}h_{70}^2~\rm M_\odot~Mpc^{-3}$, and to a Chabrier
IMF when necessary.}\label{metal_densities}
\begin{tabular}{@{}|ccc|@{}}
\hline
\multicolumn{3}{|c|}{Stellar metallicity density}\\
\hline
$\rho_Z (10^6~h_{70}~\rm M_\odot~Mpc^{-3})$ & $\Omega_Z (10^{-5}h_{70}^{-1})$ & Reference\\
\hline
$7.099^{+2.184}_{-1.943}$ & $5.219^{+1.606}_{-1.428}$ & This work \\ 
$8.40^{+6.04}_{-1.96}$ & $6.18^{+4.44}_{-1.44}$ & 1 \\
$8.96$ & $6.58$ & 2 \\
$4.62\pm1.08$ & $3.4\pm0.8$ & 3 \\
\hline
$6.57$ & $4.83$ & 4 \\
$3.4$ & $2.5$ & 5 \\
$4.2-6.4$ & $3.09-4.7$ & 6 \\
$8.7-13.3$ & $6.3-9.7$ & 7\\
\noalign{\smallskip}
\hline
\end{tabular}
\begin{minipage}{8.5truecm}
\footnotesize{Observational estimates of metal mass density in stars are from: (1)\cite{pagel02},
(2)\cite{dunne03}, (3)\cite{FP04}. We report also values of $\rho_Z$ predicted from chemo-photometric
models of \cite{cm2004} (4), from cosmic hydrodynamic simulations of \cite{DO07} (5), and by integrating
the SFH of \cite{pei99} cosmic chemical evolution models (6) and the analytic fit of \cite{cole2001} to
the dust-corrected cosmic star formation history ($\dot{\rho}_\ast$) data of \cite{steidel99} (7),
assuming $\dot{\rho}_Z=y \dot{\rho}_\ast$ (in both cases the two values correspond to $y=0.015$ and
$y=0.023$ respectively).}
\end{minipage}
\end{table}

\begin{table}
\centering
\caption{Present-day stellar mass density as derived in this work and in the
literature. All values adjusted as described in Table~\ref{metal_densities}.}\label{mass_densities}
\begin{tabular}{@{}|ccc|@{}}
\hline
\multicolumn{3}{|c|}{Stellar mass density}\\
\hline
$\rho_\ast (10^8~h_{70}~\rm M_\odot~Mpc^{-3})$ & $\Omega_\ast (10^{-3}~h_{70}^{-1})$ & Reference \\
\hline
$3.413^{+0.569}_{-0.554}$ & $2.509^{+0.418}_{-0.407}$ & This work \\ 
\noalign{\smallskip}
$3.21\pm0.47$ & $2.36\pm0.35$ & 1 \\ 
\noalign{\smallskip}
$3.78\pm0.44$ & $2.78\pm0.32$ & 2 \\
\noalign{\smallskip}
$4.21^{+3.02}_{-0.98}$ & $3.09^{+2.22}_{-0.72}$ & 3\\
\noalign{\smallskip}
$1.94-4.27$ & $1.43-3.14$ & 4 \\
\noalign{\smallskip}
$3.14\pm0.94$ & $2.3\pm0.7$ & 5\\ 
\noalign{\smallskip}
$1.76^{+0.18}_{-0.19}$ & $1.29\pm0.14$ & 6 \\ 
\noalign{\smallskip}
$2.22\pm0.2$ & $1.63\pm0.15$ & 7 \\
\noalign{\smallskip}
$2.87^{+0.74}_{-0.35}$ & $2.11^{+0.54}_{-0.26}$ & 8\\
\noalign{\smallskip}
$3.50\pm0.17$ & $2.57\pm0.13$ & 9 \\
\noalign{\smallskip}
$4.30\pm0.56$ & $3.16\pm0.41$ & 10 \\
\noalign{\smallskip}
$4.89\pm0.95$ & $3.6\pm0.7$ & 11 \\
\hline
$2.80$ & $2.06$ & 12 \\
$2.05$ & $1.51$ & 13 \\
$2.38$ & $1.75$ & 14 \\
$2.86-3.81$ & $2.1-2.8$ & 15 \\
$1.90-5.81$ & $1.40-4.27$ & 16 \\
\noalign{\smallskip}
\hline
\end{tabular}
\begin{minipage}{8.5truecm}
\footnotesize{Observational estimates of stellar mass density are from: (1) \cite{cole2001},
(2) \cite{kochanek01}, (3) \cite{fhp98} from the compilation of \cite{pagel02}, (4) \cite{glazebrook03},
(5) \cite{bell03}, (6) \cite{rudnick03}, (7) \cite{rudnick06}, (8) \cite{ben06}, (9) \cite{driver07a},
(10) \cite{driver07b}, (11) \cite{shankar06}. Model predictions are from (12) \cite{pei99}, (13)
\cite{cm2004}, (14)
\cite{monaco07}, (15) \cite{nagamine06} and integration of the analytic fit of \cite{cole2001} to
$\dot{\rho}_\ast$ data of \cite{steidel99} (16, the two values are without and with correction for dust
extinction, respectively).}
\end{minipage}
\end{table}

\section{The distribution of metals and baryons in the local Universe}\label{metal_distribution}
We now study how metals and baryons are distributed according to different galaxy properties
(Section~\ref{distrz}) and comment on the characteristic age of the mass and metallicity distributions as
a function of stellar mass (Section~\ref{age}). Finally, we compare the observed distributions in detail
with those predicted by the galaxy formation model of \cite{dLB07} applied to the Millennium Simulation
\citep[][Section~\ref{millennium}]{millennium}. 

\subsection{An inventory of the stellar metallicity and stellar mass}\label{distrz}
\begin{figure*}
\centerline{\includegraphics[width=12truecm]{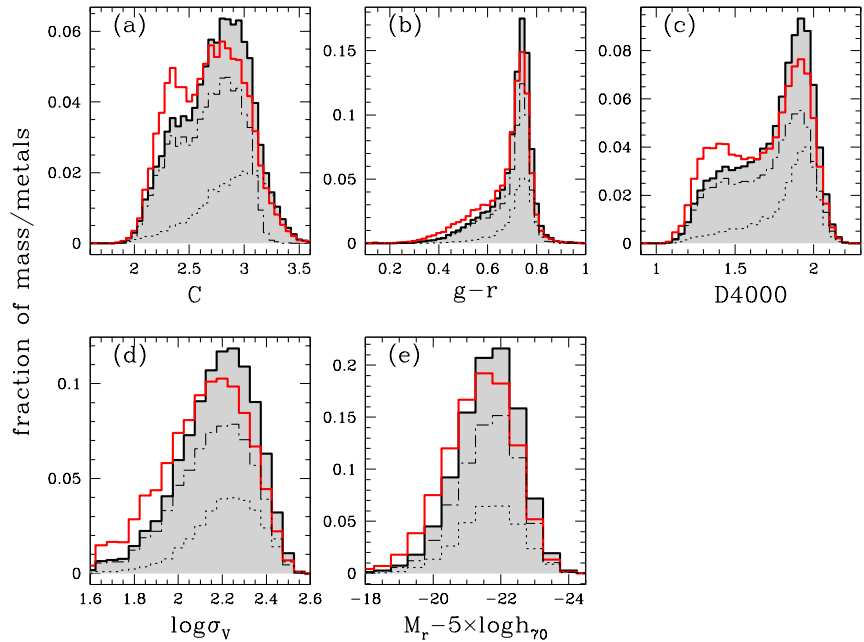}}
\caption{The fraction of the total mass of metals locked up in stars in the local Universe is shown as a
function of various observable galaxy properties: (a) the concentration parameter, (b) the rest-frame
$g-r$ colour, (c) the 4000\AA-break index strength, (d) the stellar velocity dispersion and (e) the
absolute $r$-band magnitude. The dotted line shows the contribution from high-S/N galaxies, while the
dot-dashed line  shows the contribution from low-S/N galaxies (obtained from high-S/N stacked spectra,
see text  for details). The continuous black line (grey-shaded histogram) shows the distribution obtained when both contributions
are  taken into account. The distribution of stellar mass as a function of the various properties is
described by the red solid line in each panel. Note that the resolution in the distribution versus
velocity dispersion and absolute magnitude is limited by the width of the bins in which low-S/N galaxies
are grouped to obtain the co-added spectra (0.05~dex and 0.5~mag respectively).}\label{distr_metals_obs}
\end{figure*}

\begin{figure*}
\centerline{\includegraphics[width=11truecm]{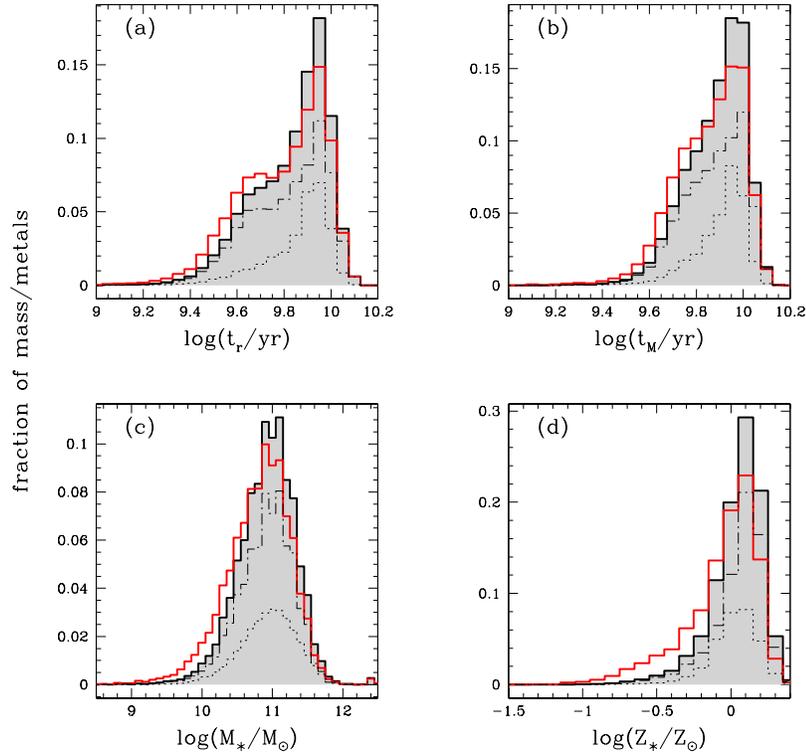}} \caption{Same as
Fig.~\ref{distr_metals_obs}, but here the fraction of stellar mass and the fraction of metals in stars
are shown as a function of the derived physical parameters: (a) $r$-band luminosity weighted age, (b)
mass-weighted age (both ages are corrected to $z=0$ by the lookback time), (c) stellar mass and (d)
stellar metallicity.}\label{distr_metals_phys}
\end{figure*}

In addition to quantifying the total metal budget in stars of the local Universe, it is of interest to
investigate the fractional contribution to the total amount of metals and baryons in stars today by
galaxies with different properties. Which are the galaxies that contain the bulk of the metals, and how
do they differ from the galaxies that contain the bulk of the stellar mass in the local Universe? In
order to answer these questions we plot in Figs.~\ref{distr_metals_obs} and~\ref{distr_metals_phys} the
fraction of the total mass of metals in stars\footnote{The differential $\rho_Z$ normalized by the total
density of the whole galaxy sample.} as a function of various galaxy properties. In
Fig.~\ref{distr_metals_obs} we analyse observable quantities, such as the concentration parameter (a),
the rest-frame $g-r$ colour (b), the 4000\AA-break index strength (c), the stellar velocity dispersion
(d) and the absolute $r$-band magnitude (e). The distribution as a function of the derived physical
properties is shown in Fig.~\ref{distr_metals_phys}: $r$-band light-weighted age (a), mass-weighted age
(b), stellar mass (c) and stellar metallicity (d). The grey-shaded histogram gives the distribution for
the sample as a whole, while the dotted and the dot-dashed lines represent the contribution from the
high-S/N galaxies only and from the low-S/N galaxies only (as derived from the stacked spectra),
respectively. It is evident that, neglecting low-S/N galaxies, we would have missed a substantial fraction
of the total amount of metals in the local Universe, in particular at low velocity dispersions, low
concentrations, low \dn\ values and hence young ages, while there is no strong segregation in luminosity
and stellar mass.

The red solid line in each panel traces for comparison the fraction of the total stellar mass as a
function of the different parameters. The stellar mass density distribution for SDSS galaxies has been
studied as a function  of spectral and photometric properties of galaxies, of their size and morphology,
stellar mass and surface  mass density by \cite{kauf03a} and \cite{jarle03}.  Our distributions
agree with those previously derived, although some differences may be expected due  to the different
sample definition. In particular, notice that the stellar mass distribution as a function of the
concentration parameter (Fig.~\ref{distr_metals_obs}a) is strongly double-peaked, whereas the
distribution shown by \cite{kauf03a} and \cite{jarle03} does not peak at any particular value. This is
likely an artifact in our distribution caused by the definition of average concentration for the coadded
spectra (see Fig.~\ref{distr_obs}c). In addition to the parameters already studied we are able to show
here the distribution of stellar mass directly as a function of age and not only of \dn.

Thanks to the good statistics provided by the SDSS DR2 we can give accurate description of the
distributions shown in Figs.~\ref{distr_metals_obs} and~\ref{distr_metals_phys}. For the velocity
dispersion $\log\sigv$ and the absolute $r$-band magnitude we are limited by the size of the bins in
which low-S/N galaxies are grouped to obtain high-S/N coadded spectra (0.05~dex and 0.5~mag
respectively). In Table~\ref{metal_frac} we give the mode of each distribution, which indicates the
typical parameter of the galaxies where metals are most likely found (last column). The
5,~10,~25,~50,~75,~90,~95 percentiles of each distribution are listed as well. The same quantities are
given also for the distribution in stellar mass density. In the last row of Table~\ref{metal_frac} we
indicate the fraction of the total stellar mass contained in galaxies that contribute different 
fractions of the total metal content. The systematic uncertainties have been estimated by calculating the
distributions with the masses, metallicities and ages corrected following Tables~\ref{sys_err_1a}
and~\ref{sys_err_1b}. The uncertainties quoted here give the range of variation in the distributions. 

From Figs.~\ref{distr_metals_obs} and~\ref{distr_metals_phys} and Table~\ref{metal_frac} it appears that
the distribution of the metals locked up in stars does not differ substantially from the distribution of
the stellar mass. In other words, the galaxies that contribute a significant fraction of the total amount
of metals in stars today are also those that contain most of the total stellar mass. The similarity in
the two distributions is determined by the relatively narrow range (roughly two orders of magnitude) in
mass covered by the stellar mass distribution: most of the weight is concentrated in galaxies with
stellar mass around $\rm 10^{11}M_\odot$. The particular shape of the mass-metallicity relation does not
influence substantially the stellar metallicity distribution. However, the increase of metallicity with
mass causes the stellar metallicity distribution to be shifted to slightly higher values of stellar mass
(see Fig.~\ref{distr_metals_phys}c). Indeed, differences are more clearly evident in galaxies with low
concentrations, low velocity dispersions and young ages: they contain a non-negligible fraction of the
total stellar mass (almost comparable to that contained in high-concentration galaxies), because they are
more numerous \citep[see e.g.][]{jarle03}, but their stars contain a much smaller fraction of metals. We
describe these results in more detail in the following.
\\

We can characterize the properties of the typical galaxy contributing stellar mass or metals by the
median of the corresponding distributions. It is remarkable that the typical galaxy in the local Universe
appears to have global properties close to that of the Milky Way and M31. More quantitatively, it has a
velocity dispersion of $\rm \sim130~km~s^{-1}$, an absolute magnitude about 1~mag brighter than
$M^{*}_r=-20.28$ in $r$-band\footnote{From \cite{blanton_lf03}, corrected to $z=0$.}, a $g-r$ colour in
agreement with what expected from the colour-magnitude relation of elliptical galaxies, a concentration
parameter of 2.7 (characteristic of a galactic disk with a significant bulge component), \dn\ of 1.7
corresponding to a fairly old stellar age of $\sim$6~Gyr, and a typical mass of
$\sim6\times10^{10}M_\odot$.
\\

Fig.~\ref{distr_metals_obs}a shows the distribution of mass and metals as a function of the concentration
parameter. The peak of the distribution of metals is at a concentration parameter of $\sim$2.9 (therefore
it is contributed by  galaxies  that are predominantly early-type).  The fraction of metals drops quickly
in galaxies with concentration parameter below the median value of $\sim$2.7. While early-type galaxies
($C\geq2.8$) contain roughly 40 percent of the total metal budget in stars, late-type galaxies 
($C\leq2.4$) contribute less than 25 percent. The contribution of early- and late-type galaxies to the
total metal budget in stars increases to 60 and 40 percent respectively if $C=2.6$ is adopted as
threshold to separate late- and early-type galaxies \citep[following][]{strateva01}. We note that the
chemo-spectrophotometric models of \cite{cm2004} predict that, while spheroids are the largest
contributors to the total amount of metals (in different phases) in the present Universe, they also
contribute significantly to the enrichment of the IGM and the majority of metals {\it in stars} come
instead from spiral galaxies (60 percent against the 40 percent contributed by spheroids). As far as the
stellar baryon fraction is concerned, we find that the two classes of galaxies contribute the same
fraction of the stellar mass density (30 or 50 percent, depending on which of the two concentration cuts
we adopt). This is consistent with what was already found by \cite{kauf03a}.
\\

The distributions of metals and baryons are very similar to each other also as a function of colour
(Fig.~\ref{distr_metals_obs}b). Red galaxies contribute the same fraction to metals as to baryons in
stars. The metal fraction becomes smaller than the stellar mass fraction only in galaxies bluer than
$g-r=0.5$. At least half and up to 75 percent of the total stellar mass (and metals) are contained in
red-sequence galaxies (assuming $g-r=0.7$ or 0.6 as colour cut respectively). This result is in good
agreement with what was found by \cite{bell03} separating elliptical galaxies with a magnitude-dependent
colour cut, and by \cite{baldry04} who also distinguish between red-peak galaxies according to their
distribution in the colour-magnitude plane. We note that these fractions are also consistent with the
distributions as a function of concentration discussed above, given that 84 percent of $C\geq2.6$
galaxies satisfy also the colour-based selection of \cite{bell03}.  
\\

The distribution of stellar metallicity as a function of \dn\ (Fig.~\ref{distr_metals_obs}c) deviates
from the distribution of stellar mass in the range of \dn\ occupied by late-type, star-forming galaxies.
Both distributions show a strong peak at \dn=1.9: galaxies with \dn\ above this value contain roughly 25
percent of the metals and 25 percent of the mass in stars today. The distribution in mass is clearly
bimodal and has a secondary peak at $\dn\sim1.4$. Galaxies with $\dn\leq1.4$
contribute another 25 percent to the total stellar mass, but only 10 percent of the metals.
\\

Fig.~\ref{distr_metals_phys} complements the picture derived from observational properties with physical
ones. Panels (a) and (b) illustrate that the differences in the stellar metallicity and stellar mass
distributions with respect to \dn\ and colour are reflected in the stellar age. Focusing on the $r$-band
light-weighted age (Fig.~\ref{distr_metals_phys}a), galaxies older than 8.5~Gyr contribute the same
fraction (25 percent) of the total stellar mass and the total stellar metallicity densities in the local
Universe, and only 5 percent comes from galaxies older than 10~Gyr. The distribution in stellar
metallicity declines rapidly at ages younger than 6.3~Gyr ($\log(t_r/yr)\sim9.8$), where roughly 50
percent of the total stellar mass, but less than 40 percent of the total amount of metals, comes from.
Similar results are found considering the mass-weighted age (Fig.~\ref{distr_metals_phys}b). The only
significant difference is that the distributions here are narrower due to the older mass-weighted age
in young, low-mass galaxies with respect to their light-weighted age. 
\\

The dependence of the fraction of mass and metals in stars on stellar mass is shown in
Fig.~\ref{distr_metals_phys}c, quantifying what expected on the basis of the distributions against
velocity dispersion and absolute magnitude, both tracers of the total stellar mass, especially in
quiescent elliptical galaxies (Fig.~\ref{distr_metals_obs}d,e).  Half of all the metals locked up in
stars today are contained in galaxies more massive than $\rm 7.2\times 10^{10}M_\odot$ (or with velocity
dispersion higher than $\rm 148~km~s^{-1}$), which contain roughly 40 percent of the total stellar mass.
Galaxies with masses below $\rm 4\times10^{10}M_\odot$ (or $\rm \sigv \la 110~km~s^{-1}$)  contain only 25
percent of the total metal budget and about 35 percent of the total stellar mass. Similarly, panel (d)
shows the (mass- and volume-weighted) projection of the mass-metallicity relation onto the metallicity
axis. This illustrates, consistently, that at least half of the metals are contained in galaxies with
metallicity above solar, which are predominantly massive ellipticals and the bulges of massive late-types,
galaxies with masses above $\rm 10^{11}M_\odot$. The steepening of the mass-metallicity relation becomes
clear at metallicities below $\rm 0.7\times Z_\odot$ (or below the transition mass of $\rm
10^{10.5}M_\odot$), where the largest differences in the relative contribution to the amount of baryons
and to the amount of metals are seen.
\\

In conclusion, we find that the bulk of the total metals locked up in stars in the local Universe resides
in galaxies with masses just above the transition mass in the mass-metallicity relation, with
morphology and spectral properties of intermediate-type galaxies (early late-types or ellipticals), and 
with fairly old stellar populations.
Given the shape of the mass density distribution, these results are in 
agreement with the correlations between stellar metallicity, age and stellar mass studied in paper~I.
Nonetheless, late-type, star-forming galaxies (with masses below the characteristic mass of $\rm
3\times10^{10}M_\odot$, low D4000 values and concentration parameters characteristic of disc-dominated
galaxies) contribute roughly 20 percent of the total mass density of metals in stars and a
slightly higher fraction of the total stellar mass density ($30-35$ percent). 


\begin{table*}
\centering
\caption{For each parameter X, the percentiles of the distribution of stellar mass and of metals as a
function of X are given. In the last column we indicate the mode of each distribution. The last row
quotes the fraction of the total stellar mass in the local Universe contained in  galaxies that
contribute different fractions of the metal content.}\label{metal_frac}
\begin{minipage}{15truecm}
\hspace{-2.truecm}
\begin{tabular}{@{}|lrrrrrrrr|@{}}
\hline
\multicolumn{9}{|c|}{Stellar mass distribution}\\
\hline
Parameter& 5\% & 10\% & 25\% & 50\% & 75\% & 90\% & 95\% & mode \\
\hline
$\log\sigma_V$  	 &$1.75_{-0.03}^{0.01}$&$1.83_{-0.03}^{+0.01}$&$1.98_{-0.03}^{+0.01}$&$2.12_{-0.03}^{+0.01}$&$2.24_{-0.02}^{+0.01}$&$2.34_{-0.01}^{+0.01}$&$2.38_{-0.01}^{+0.00\dag}$&$2.20_{-0.05}^{+0.02}$\\
\noalign{\smallskip}
$M_r-5\times\log h_{70}$ &$-23.17_{-0.03}^{+0.06}$&$-22.86_{-0.03}^{+0.04}$&$-22.33_{-0.03}^{+0.06}$&$-21.65_{-0.04}^{+0.09}$&$-20.90_{-0.04}^{+0.10}$&$-20.16_{-0.03}^{+0.08}$&$-19.69_{-0.03}^{+0.08}$&$-21.50_{-0.25}^{+0.25}$\\
\noalign{\smallskip}
$C$			 &$2.18_{-0.04}^{+0.01}$&$2.26_{-0.05}^{+0.01}$&$2.42_{-0.07}^{+0.02}$&$2.69_{-0.05}^{+0.02}$&$2.91_{-0.03}^{+0.02}$&$3.07_{-0.02}^{+0.01}$&$3.17_{-0.02}^{+0.01}$&$2.80_{-0.45}^{+0.05}$\\
\noalign{\smallskip}
\dn\			 &$1.24_{-0.01}^{+0.00\dag}$&$1.30_{-0.02}^{+0.01}$&$1.45_{-0.04}^{+0.01}$&$1.73_{-0.05}^{+0.02}$&$1.89_{-0.02}^{+0.01}$&$1.97_{-0.01}^{+0.00\dag}$&$2.01_{-0.01}^{+0.00\dag}$&$1.92_{-0.02}^{+0.02}$\\
\noalign{\smallskip}
$g-r$                    &$0.44_{-0.02}^{+0.00\dag}$&$0.51_{-0.03}^{+0.01}$&$0.62_{-0.03}^{+0.01}$&$0.71_{-0.01}^{+0.00\dag}$&$0.74_{-0.00\dag}^{+0.00\dag}$&$0.77_{-0.00\dag}^{+0.00\dag}$&$0.79_{-0.00\dag}^{+0.00\dag}$&$0.74_{-0.01}^{+0.01}$\\
\noalign{\smallskip}
$\log (M_\ast/M_\odot)$  &$9.88_{-0.06}^{+0.06}$&$10.11_{-0.07}^{+0.06}$&$10.43_{-0.07}^{+0.06}$&$10.77_{-0.09}^{+0.06}$&$11.04_{-0.10}^{+0.06}$&$11.26_{-0.10}^{+0.06}$&$11.39_{-0.10}^{+0.07}$&$10.90_{-0.20}^{+0.10}$\\
\noalign{\smallskip}
$\log (Z_\ast/Z_\odot)$  &$-0.64_{-0.04}^{+0.06}$&$-0.47_{-0.05}^{+0.06}$&$-0.21_{-0.06}^{+0.05}$&$-0.03_{-0.07}^{+0.06}$&$0.09_{-0.07}^{+0.06}$&$0.17_{-0.07}^{+0.06}$&$0.20_{-0.04}^{+0.07}$&$0.10_{-0.10}^{+0.10}$\\
\noalign{\smallskip}
$\log (t_r/yr)$ 	 &$9.44_{-0.10}^{+0.08}$&$9.51_{-0.10}^{+0.09}$&$9.64_{-0.10}^{+0.08}$&$9.81_{-0.10}^{+0.06}$&$9.92_{-0.10}^{+0.07}$&$9.97_{-0.10}^{+0.09}$&$10.00_{-0.10}^{+0.10}$&$9.95_{-0.10}^{+0.05}$\\
\noalign{\smallskip}
$\log (t_M/yr)$ 	 &$9.57_{-0.09}^{+0.08}$&$9.64_{-0.09}^{+0.08}$&$9.74_{-0.09}^{+0.07}$&$9.86_{-0.09}^{+0.06}$&$9.94_{-0.09}^{+0.07}$&$9.99_{-0.09}^{+0.09}$&$10.02_{-0.09}^{+0.09}$&$9.95_{-0.05}^{+0.10}$\\
\hline
\multicolumn{9}{|c|}{Stellar metallicity distribution}\\
\hline
$\log\sigma_V$  	 &$1.84_{-0.04}^{+0.01}$&$1.93_{-0.05}^{+0.01}$&$2.05_{-0.05}^{+0.01}$&$2.17_{-0.04}^{+0.01}$&$2.28_{-0.03}^{+0.01}$&$2.36_{-0.02}^{+0.01}$&$2.40_{-0.01}^{+0.01}$&$2.25_{-0.05}^{+0.02}$\\
\noalign{\smallskip}
$M_r-5\times\log h_{70}$ &$-23.31_{-0.04}^{+0.11}$&$-22.98_{-0.05}^{+0.08}$&$-22.50_{-0.06}^{+0.12}$&$-21.91_{-0.06}^{+0.17}$&$-21.24_{-0.06}^{+0.15}$&$-20.61_{-0.05}^{+0.12}$&$-20.19_{-0.05}^{+0.13}$&$-22.00_{-0.25}^{+0.50}$\\
\noalign{\smallskip}
$C$			 &$2.21_{-0.07}^{+0.01}$&$2.31_{-0.08}^{+0.02}$&$2.53_{-0.11}^{+0.04}$&$2.77_{-0.08}^{+0.04}$&$2.96_{-0.06}^{+0.02}$&$3.10_{-0.04}^{+0.02}$&$3.20_{-0.04}^{+0.01}$&$2.95_{-0.25}^{+0.00\dag}$\\
\noalign{\smallskip}
\dn\			 &$1.28_{-0.03}^{+0.01}$&$1.36_{-0.05}^{+0.02}$&$1.57_{-0.08}^{+0.03}$&$1.80_{-0.05}^{+0.02}$&$1.91_{-0.02}^{+0.01}$&$1.99_{-0.01}^{+0.00\dag}$&$2.02_{-0.01}^{+0.00\dag}$&$1.92_{-0.02}^{+0.02}$\\
\noalign{\smallskip}
$g-r$                    &$0.50_{-0.04}^{+0.01}$&$0.57_{-0.04}^{+0.01}$&$0.67_{-0.04}^{+0.01}$&$0.72_{-0.01}^{+0.00\dag}$&$0.75_{-0.01}^{+0.00\dag}$&$0.78_{-0.01}^{+0.00\dag}$&$0.79_{-0.01}^{+0.00\dag}$&$0.74_{-0.01}^{+0.01}$\\
\noalign{\smallskip}
$\log (M_\ast/M_\odot)$  &$10.12_{-0.06}^{+0.07}$&$10.30_{-0.07}^{+0.06}$&$10.59_{-0.09}^{+0.06}$&$10.86_{-0.15}^{+0.06}$&$11.10_{-0.13}^{+0.07}$&$11.31_{-0.13}^{+0.08}$&$11.44_{-0.13}^{+0.08}$&$10.90_{-0.20}^{+0.10}$\\
\noalign{\smallskip}
$\log (Z_\ast/Z_\odot)$  &$-0.34_{-0.05}^{+0.06}$&$-0.21_{-0.07}^{+0.05}$&$-0.07_{-0.07}^{+0.06}$&$0.05_{-0.07}^{+0.06}$&$0.14_{-0.07}^{+0.05}$&$0.20_{-0.04}^{+0.07}$&$0.25_{-0.05}^{+0.05}$&$0.10_{-0.10}^{+0.10}$\\
\noalign{\smallskip}
$\log (t_r/yr)$ 	 &$9.51_{-0.10}^{+0.08}$&$9.58_{-0.10}^{+0.08}$&$9.71_{-0.10}^{+0.07}$&$9.85_{-0.10}^{+0.06}$&$9.93_{-0.10}^{+0.08}$&$9.98_{-0.10}^{+0.10}$&$10.00_{-0.10}^{+0.10}$&$9.95_{-0.10}^{+0.10}$\\
\noalign{\smallskip}
$\log (t_M/yr)$ 	 &$9.63_{-0.09}^{+0.08}$&$9.68_{-0.09}^{+0.07}$&$9.78_{-0.09}^{+0.07}$&$9.89_{-0.09}^{+0.06}$&$9.96_{-0.09}^{+0.08}$&$10.00_{-0.09}^{+0.09}$&$10.03_{-0.09}^{+0.09}$&$9.95_{-0.05}^{+0.10}$\\
\hline
\multicolumn{9}{|c|}{Fraction of the total stellar mass contributed}\\
\hline
			 &0.106 & 0.178 & 0.351 & 0.592 & 0.806 & 0.925 &  0.962 & \\
\hline
\end{tabular}
\footnotesize{\dag The estimated systematic uncertainty is smaller than 0.01, i.e. less than 20 percent
of the bin size.}
\end{minipage}
\end{table*}

\subsection{The characteristic age of the mass and metallicity distributions}\label{age}
We investigate further the distribution of metals and baryons as a function of stellar age. As shown in
Fig.~\ref{distr_metals_phys}a,b both the stellar mass distribution and the metals distribution have a
peak at a mean age of $\log(t/yr)=9.95$ (almost 9~Gyr). If we could translate this characteristic age
into a redshift, it would correspond to a characteristic formation redshift of $z\sim1.4$, which nicely
falls in the redshift range over which the cosmic metal production rate and the cosmic star formation
rate are expected to decline \citep[e.g.][and references
therein]{lilly96,madau96,madau98,glazebrook04,hopkins06}. This is of course only a naive interpretation,
because we can only assign an average age (with a larger weight to younger stars) to individual galaxies
and we cannot describe the real distribution of stellar ages within a galaxy. This is even more critical
when the distribution in stellar metallicity is considered, as long as we assign a fixed metallicity to
all stars in a galaxy, rather than following the chemical evolution along the star formation history. It
is however interesting to notice that the result obtained from this `archaeological' approach is close to
what expected from `direct' investigation of the cosmic star formation history.

\begin{figure*}
\centerline{\includegraphics[width=7truecm]{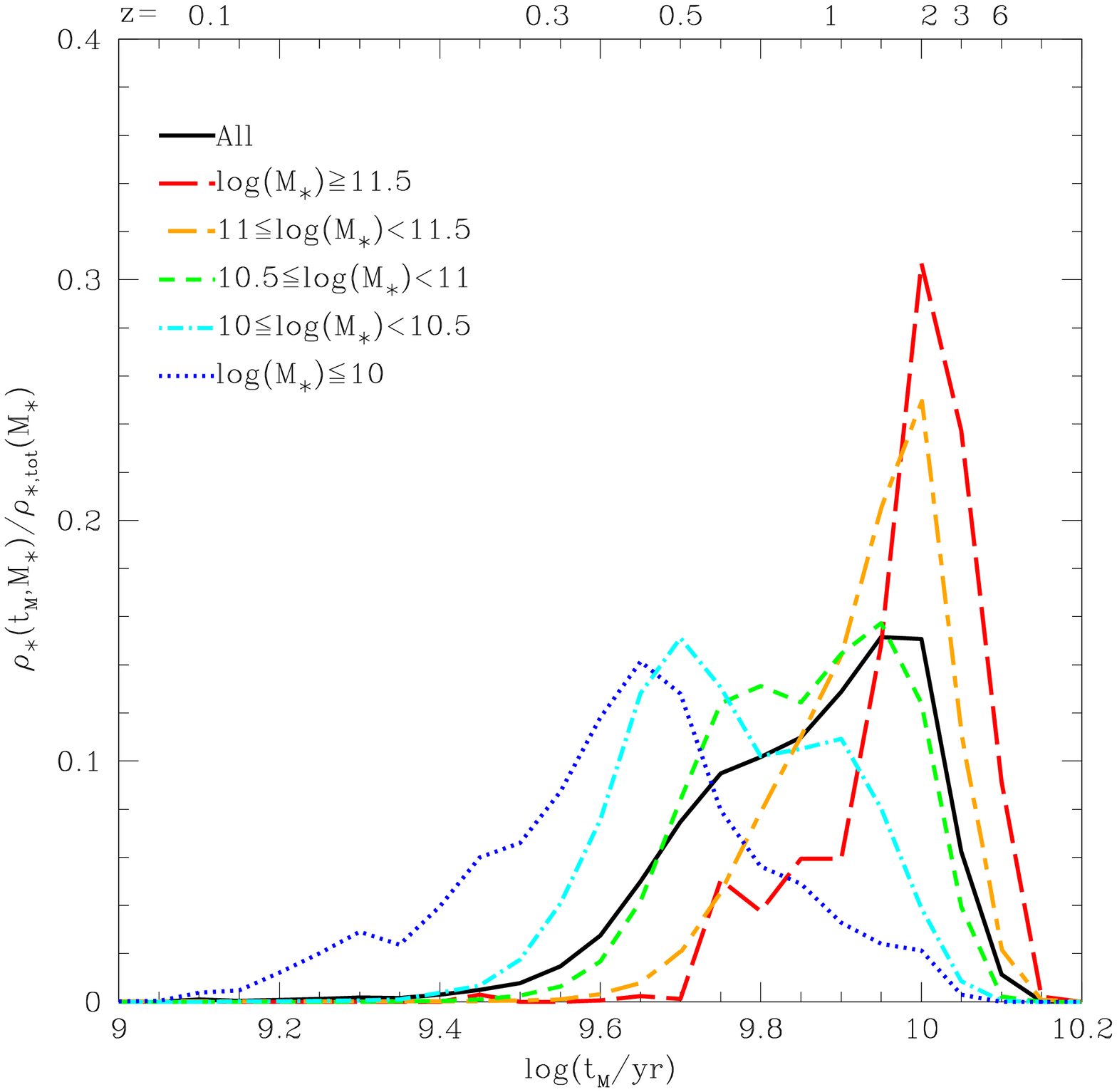}
\includegraphics[width=7truecm]{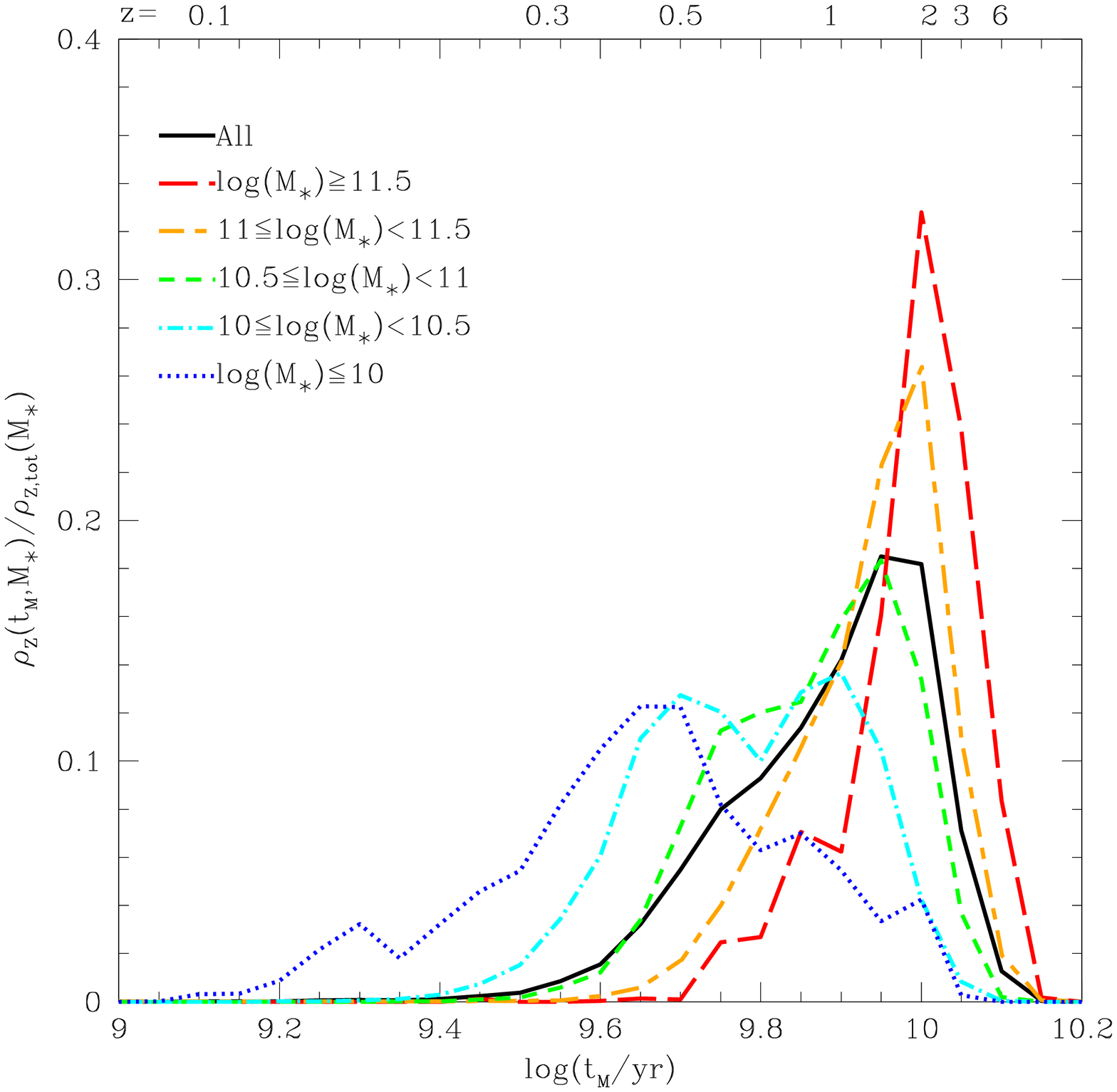}}
\caption{Differential distribution of stellar mass density (left panel) and mass density of metals in
stars (right panels) as a function of mass-weighted age for the whole sample (continuous line). Different
line styles distinguish the distribution for galaxies in different mass bins. Each distribution is
normalized to the total $\rho_\ast$ and $\rho_Z$ in the corresponding mass bin. For reference the upper
x-axis gives the redshift corresponding to a given mean age.}\label{distr_mbin}
\end{figure*}

Moreover, there is evidence that the timescale of star formation depends on the mass of the galaxy, with
more massive galaxies having an early and shorter star formation, and less massive galaxies having a star
formation more extended toward the present day. If so, the distribution of stellar ages in individual
massive galaxies would be narrower than in less massive galaxies around the (mass- or light-weighted)
average age, which would hence be more representative of the average formation redshift of the stars. It is thus
useful to look at the distributions of baryon and metal densities in stars as a function of stellar age for
galaxies with similar masses. These are shown in Fig.~\ref{distr_mbin} (left- and right-hand panel
respectively). The black continuous line shows the distribution of $\rho_\ast$ and $\rho_Z$
for the sample as a whole, while different colours and line styles correspond to different stellar mass
bins (each distribution is normalized to the total $\rho_\ast$ and $\rho_Z$ in the corresponding mass
bin). For reference, we indicate in the upper x-axis the redshift corresponding to the age on the lower
x-axis (interpreted as lookback time from the present).

The distribution for the entire galaxy population is traced very well by the distribution of galaxies in
the mass range between $3\times 10^{10}$ and $\rm 10^{11} M_\odot$, which alone contain 38 percent of the
total stellar mass budget (these galaxies also provide the largest contribution to the star formation
rate density and stellar mass density up to $z\sim 1$, as shown e.g. by \cite{ben06,borch06,xxz07}). This
is true for both the stellar mass density and the metal density in stars. The contribution to the total
baryon and metal budget in stars from galaxies with old stellar populations increases significantly from
the lowest- to the highest-mass bin. This reflects the median relation between the galaxy mean age and
stellar mass (see e.g. fig.~8 of paper~I). Not only do the distributions in mass-weighted age shift
gradually to younger ages as less massive galaxies are considered, but they also span larger ranges in
age\footnote{It is interesting to notice that the bimodal distribution expected from the bimodality in
\dn\ appears in galaxies of intermediate masses. This would be even clearer if we plotted light-weighted
age instead of mass-weighted age. The bimodality in age is instead almost lost when the full population
is considered.}, a result of the increasing scatter in age at lower masses in the age-mass
relationship.\footnote{Note that these are logarithmic ranges, which means that it is the relative age
range, not the absolute one, which increases on a linear scale toward lower masses.} 

\begin{figure}
\centerline{\includegraphics[width=9truecm]{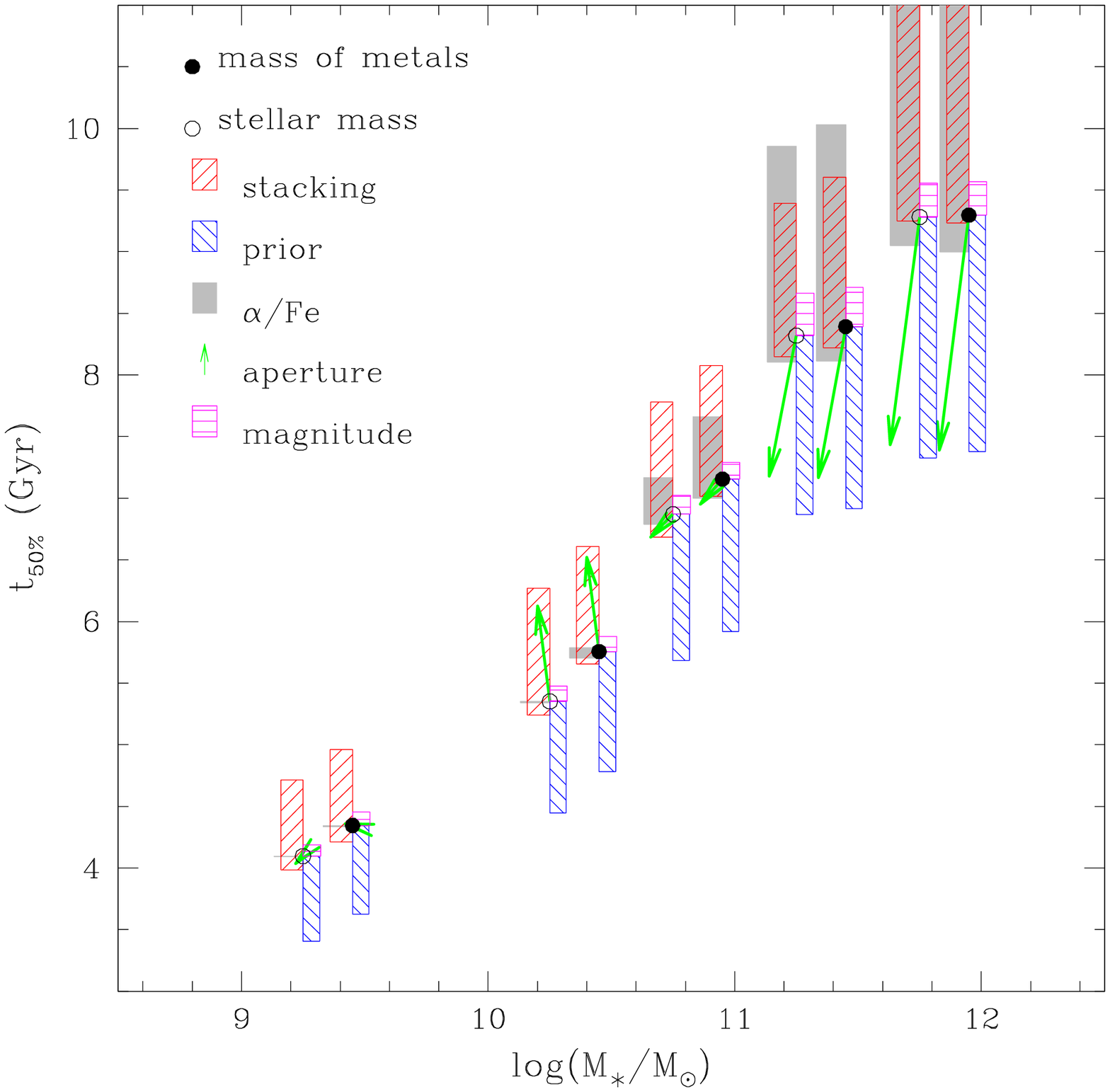}}
\caption{Mass-weighted age above which half of the total amount of mass (empty circles) and metals
(filled circles) in stars are contained at the present epoch, as a function of stellar mass. Galaxies
have been divided in the same five bins of stellar mass as in Fig.~\ref{distr_mbin}. Boxes and arrows
quantify the ranges of the different sources of systematic uncertainties.}\label{t50_mass}
\end{figure}

Quantitatively, 25 percent of the total stellar mass and metal density of the most massive galaxies ($\rm
>10^{11}M_\odot$) is contained in galaxies older than 9.5 Gyr, and half of it in galaxies older than 8.5
Gyr (corresponding to redshift greater than 1.2). As much as 75 percent of the stellar mass and metal
density of low-mass galaxies (below $\rm 10^{10}M_\odot$) is distributed in galaxies with mass-weighted
ages younger than $\sim$5.5 Gyr, and 50 percent in galaxies younger than $\sim$4.5 Gyr (corresponding to
redshifts below $\sim$0.5). This trend is illustrated in Fig.~\ref{t50_mass}. In each of the stellar mass
bins defined in Fig.~\ref{distr_mbin}, we define two characteristic mass-weighted ages as the median of
the distribution of stellar mass and the distribution of metals versus age. These characteristic ages are
plotted as a function of stellar mass in Fig.~\ref{t50_mass}. We also indicate the effect of individual
sources of systematic uncertainties in each mass bin. The aperture bias and the scaled-solar abundance
ratio of the models, in particular, affect to a larger extent more massive galaxies. 

Despite the large uncertainties, it is clear from Fig.~\ref{t50_mass} that the characteristic ages of the
stellar mass and metallicity distributions become progressively younger in less massive galaxies. This
is in agreement (at least qualitatively) with the findings that the characteristic epoch at which the
bulk of the stars (and metals) in present-day massive (mostly elliptical) galaxies formed is at redshift
at least greater than 1 \citep[e.g.][and references therein]{thomas05}. Finally, it is worth noting that
the characteristic age of the distribution of metals is never younger than the characteristic age of the
distribution of mass. This is particularly true at masses below $\rm 10^{11}M_\odot$. This probably
indicates that the correlation between age and stellar metallicity persists in individual stellar mass
bins, with older galaxies being also more metal-rich. This correlation is weaker in higher-mass galaxies,
because of the narrower ranges in both age and metallicity and because contrasted by the age-metallicity
degeneracy (see e.g. figs. 11 and 12 of paper~I).

\subsection{Comparison with the Millennium Simulation}\label{millennium}
The distribution functions of mass density of baryons and metals in stars provide important quantitative
constraints at redshift zero against which models of galaxy formation and evolution in a cosmological
context should be tested. We provide here a first comparison between our observational results and model
predictions. We consider the results of the {\it Millennium Run}, the largest N-body simulation of
structure formation carried out so far within the $\Lambda$CDM cosmological paradigm \citep{millennium}. 
We make use of the galaxy catalogues produced by semi-analytic galaxy formation algorithms run on the
Millennium merger trees, as described in \cite{croton06} and \cite{dLB07}, and select galaxies at $z=0$.

Before going into the detail of the distributions of metals and baryons in stars, we calculate the
total $\rho_\ast$ and $\rho_Z$ predicted by the simulation. The total stellar mass
density is $\rm 3.203\times10^8~M_\odot~Mpc^{-3}$, in agreement with our determination as already
mentioned in Section~\ref{literature}. The total mass density of metals in stars is instead $\rm
4.053\times10^6~M_\odot~Mpc^{-3}$, lower than what observed (in particular it is $1.5\sigma$ lower than
our result). This originates from the narrow metallicity distribution predicted by the models and the
lower metallicity for massive galaxies, as we discuss below. 

It is also of interest to mention the metal and baryon mass densities in other gaseous components, inside
or outside galaxies, as predicted by the Millennium Simulation. The mass and metal densities in stars represent the 17.6 and 25.3 percent of the
corresponding total values. Similar contributions (13.5 and 11.8 percent for the baryons and the metals,
respectively) come from the gas ejected from galaxies into the surrounding intergalactic medium 
\citep[see][]{delucia04a}. The cold gas inside galaxies provides only 6.6 percent and 9.8 percent to the
total density of baryons and metals, respectively. The remaining 62.2 and 53.1 percent is contained in the
hot gas component. The baryon fractions derived from the Millennium Simulation are in agreement with
those predicted by the chemo-photometric models of \cite{cm2004}. In both cases the fraction of baryons
in stars is higher than what predicted by the simulations of \cite{DO07} and than what we observe. The
mass fraction of metals in stars predicted by the Millennium Simulation is instead about half of what
quoted by \cite{cm2004} and \cite{DO07}.

\begin{figure*}
\centerline{\includegraphics[width=10truecm]{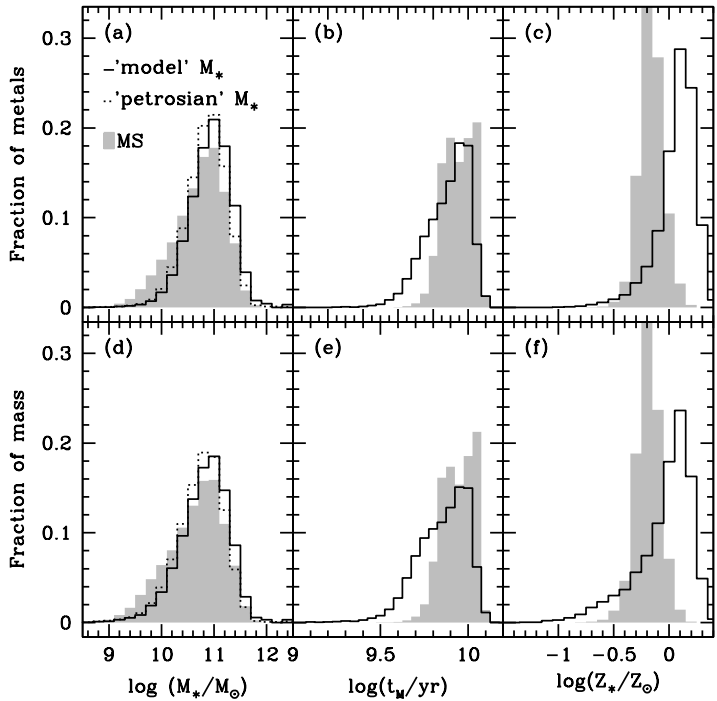}}
\caption{Comparison between predicted and observed distributions of stellar mass and metals. The fraction
of metals in stars (upper panels) and the fraction of mass in stars (lower panels) are shown as functions
of stellar mass (a,d), mass-weighted age (b,e), stellar metallicity (c,f). The shaded histograms show the
distributions derived using the semi-analytic galaxy catalogue of the Millennium Simulation. The
continuous line shows the results from our analysis on SDSS. The dotted histogram in panel (a) and (d)
shows the results obtained when normalizing the mass-to-light ratio by the $z$-band Petrosian magnitude
instead of model magnitude to derive stellar mass.}\label{MS}
\end{figure*}

We now discuss the contribution to the mass and metals in stars by different galaxies. After
selecting galaxies at $z=0$ from the Millennium Catalogue, we construct the stellar mass-weighted
distributions of baryons and metals in stars as functions of the galaxy stellar mass, mass-weighted age
and stellar metallicity. Fig.~\ref{MS} compares such distributions (gray-shaded histograms) with those
derived from our SDSS measurements (solid lines).
 
Panels (a) and (d)  of Fig.~\ref{MS} illustrate the fraction of mass and metals as a function of stellar
mass. The distributions predicted by the models display an offset in stellar mass with respect to the
observationally derived distributions of $\sim$0.06~dex. We note that this offset is reduced when we use
stellar masses based on Petrosian instead of model magnitudes (as shown by the dotted line). Comparison
between the shaded and dotted histograms in panels (a) and (d) shows a good agreement between the
observational results and the predictions from the simulations. It is clear, however, that the models
predict too much mass in low-mass galaxies with respect to observations. This feature of the models has
been already observed in the comparison with observational data of the $b_J$- and K-band luminosity
function and the stellar mass function \citep[see e.g.][]{croton06,bertone07}. Recently, \cite{bertone07}
have presented a new implementation of a feedback scheme which incorporates a dynamical treatment of
galactic winds powered by supernovae and by stellar winds, as alternative to empirically-motivated
schemes of supernovae feedback, but without replacing AGN feedback. This treatment seems to alleviate the
discrepancy in the galaxy stellar mass function, by reducing the abundance of low-mass galaxies and
reaching a better agreement, in this respect, with observations.

Panels (b) and (e) show the fraction of baryons and metals in stars as a function of the galaxy
mass-weighted age. While the predicted and observed distributions agree reasonably well at the oldest
ages, there is clearly a deficit of young galaxies (with ages younger than $\log t=9.8$) in the
simulations. This discrepancy could originate, at least in part, from the fact that, although we use
mass-weighted ages for this comparison, the observationally derived mean ages are always extracted from
fits to the galaxy spectra, and hence are always somehow weighted by light. This could thus give more
weight to the youngest stellar populations in the galaxy. We note however that the tendency of the models
to produce too old stellar populations with respect to those that we derive appears also in the relation
between the {\it light}-weighted age and the stellar mass of elliptical galaxies (as seen in the comparison
of fig.~6a of \cite{delucia05} and fig.~17d of paper~II), even though there is a good qualitative
agreement.

Finally, panels (c) and (f) compares the predicted and observed distribution as a function of stellar
metallicity. The disagreement between the observational result and model prediction is caused by the
narrow stellar metallicity range spanned by the simulations. Moreover, while the observational trend of
increasing metallicity with increasing mass is reproduced by the models, the predicted relation is too
shallow with respect to observations, hardly reaching solar metallicity for the most massive galaxies.
The galactic wind feedback scheme implemented by \cite{bertone07} seems to provide a better agreement
with observations, causing a steepening of the mass-metallicity relation (due to the high efficiency of
mass ejection in small haloes which suppresses star formation and hence the amount of metals that can be
produced and locked up in stars) and predicting super-solar metallicities in massive galaxies. 

As a general remark it is also interesting to notice the similarity between the distributions of mass and
the corresponding distributions of metals in stars. While this is true also for the observationally
derived distributions (as discussed in Section~\ref{distrz}), it appears to a greater extent in the
models. This originates from the narrower range in stellar metallicity and the shallower stellar
metallicity/stellar mass relation with respect to observations.

\section{Summary and conclusions}\label{conclusions}
In this work we have exploited recent estimates of physical parameters, such as stellar metallicity and
stellar mass, for a comprehensive sample of more than $10^5$ nearby galaxies to derive the total mass
density of metals and baryons locked up in stars in the local Universe, also expressed in terms of the
present-day mass-weighted average stellar metallicity. Moreover, we have quantified the
contribution to the total amount of metals by galaxies with different morphological and spectral
properties, and compared the distribution of the stellar metallicity density with that of the stellar
mass density.

The sample used is drawn from the SDSS DR2 and it includes galaxies spanning a wide range of star
formation activity, from quiescent early-type to actively star-forming galaxies. The stellar
metallicities, ages and stellar masses of these galaxies were previously derived
(paper~I) by comparing an optimally selected set of spectral absorption features to a
comprehensive Monte Carlo library of model star formation histories (SFH), based on the high-resolution
\cite{bc03} population synthesis code. We have shown in paper~I that the uncertainties on the
derived physical parameters directly depend on the mean spectral signal-to-noise ratio (S/N). Because of
this, in previous work we have focused on galaxies with S/N greater than 20, biasing our sample
towards more concentrated, higher surface brightness and higher velocity dispersion galaxies. 

We need to include here also galaxies at lower S/N in order to derive a fair estimate of the total
metallicity of stars at $z=0.1$. We do this by coadding (weighting by 1/\vmax) the spectra of
low-S/N galaxies with similar $r$-band absolute magnitude, velocity dispersion and \dn\ until a minimum
S/N requirement is satisfied. This allows us to derive estimates of the (1/\vmax-weighted average)
stellar metallicity of low-S/N galaxies with an uncertainty not greater than 0.2~dex, i.e. comparable to
the accuracy with which stellar metallicity is derived from individual high-S/N spectra. A similar
improvement is obtained for light- and mass-weighted age and stellar mass, although these parameters are
less affected than stellar metallicity by the quality of the spectrum. 

We estimated the total mass density of baryons and of metals locked up in stars in the local Universe, by
combining the contribution of the individual high-S/N galaxies and the low-S/N galaxies, as derived from
the coadded spectra. We find, respectively,
$\rho_\ast=3.413\pm0.005^{+0.569}_{-0.554}\times10^8~h_{70}~\rm M_\odot~Mpc^{-3}$ and
$\rho_Z=7.099\pm0.019^{+2.184}_{-1.943}\times10^6~h_{70}~\rm M_\odot~Mpc^{-3}$. This is in broad
agreement with other measures from the literature, which however span a relatively large range. The
measure of the metal mass density is perhaps more significant in discriminating different model
predictions or approaches. We find a reasonably good agreement with results of chemical evolution models,
such as those of \cite{pei99} or \cite{cm2004}. Hydrodynamic simulations seem to predict values of
$\rho_Z$ at the lower end of the observational constraints. The total amount of metals locked up in stars
is sensitive to the feedback efficiency, which regulates the amount of gas and metals expelled from
galaxies. Integration of cosmic star formation histories, on the other hand, in general overpredicts the
total amount of metals in stars. This is probably due to poor knowledge of dust corrections at high
redshift and uncertainties on the average metal yield to convert star formation rate into metal
production rate.
 
By combining the densities of mass and metals in stars we can estimate the average stellar metallicity
today to be consistent with solar (Equation~\ref{eqn6}). This is not surprising, given that the stellar
mass budget is dominated, at least in the local Universe, by galaxies just above the transition mass
(approximately $L^\ast$ galaxies), which have typically solar metallicity. This value is in agreement
with the prediction of the chemo-photometric models of \cite{cm2004} and the result of \cite{edmunds97}. 

The good statistics available from the SDSS allows us to provide an analysis of the distribution of the
stellar metallicity density as a function of global galaxy properties (such as velocity dispersion,
luminosity and stellar mass), as a function of morphology (as approximated by the concentration
parameter), and as a function of spectral and physical properties of the stellar populations (such as
colour and 4000\AA-break strength, stellar age and metallicity). We have compared such distributions with
the corresponding distributions of stellar mass density. Our study has shown that the stellar
metallicity and the stellar mass distributions do not differ significantly, in particular at the
high-mass end. In other words, the galaxies that contribute most of the total amount of metals in stars
have properties similar to those containing the bulk of the total mass in stars.

The typical galaxy in the present-day Universe where baryons and metals reside has a concentration
parameter of 2.7 (indicating a disc and a significant bulge component), a $\dn=1.7$, corresponding to a
fairly old stellar age of $\sim$6~Gyr, and a stellar mass of $6\times10^{10}M_\odot$. Note that the
characteristic mass found here is above the stellar mass of $3\times10^{10}M_\odot$ at which the
transition from low-mass, metal-poor, disc-dominated galaxies to high-mass, metal-rich, bulge-dominated
galaxies occurs. Galaxies with masses greater than $3\times10^{10}M_\odot$, where the mass-metallicity
relation starts to flatten \citep{christy04,paperI}, contain more than 80 percent of the total amount of
mass and metals in stars.

The stellar metallicity and stellar mass distributions differ most significantly with respect to
concentration and stellar age (as indicated also by the colour and \dn), in the sense that the relative
contributions to the total stellar mass and to the total metallicity are reversed in late-type galaxies,
compared to early-type galaxies (morphologically or photometrically classified as such). Separating
galaxies on the basis of their concentration parameter, we find that bulge-dominated and disc-dominated
galaxies provide the same fraction of the total mass in stars (30 or 50 percent depending on the cut in
$C$ adopted). On the contrary, the contribution of disc-dominated galaxies to the mass density of metals 
is lower than that of bulge-dominated galaxies (with the same cuts in concentration). For example,
defining early-type galaxies as those having $C\geq2.8$ and late-types those having $C\leq2.4$, the two
classes of galaxies both contribute 30 percent to the stellar mass budget, but 40 and less than 25
percent respectively to the total metal budget in stars.

Similar considerations hold for the distributions as functions of colour or \dn. We find that at least 50
up to 75 percent of the stellar mass (and of the metals) is locked into red-sequence galaxies (depending
on the assumed colour cut). These fractions are in good agreement with previous results from e.g.
\cite{bell03} and \cite{baldry04}. We show that these observational distributions are physically
translated into the corresponding distributions in age. More than 50 percent of the mass and the metals
locked up in stars reside in galaxies with ages older than 6.3~Gyr.

It would be tempting to translate the characteristic age of the stellar populations contributing most of
the metals and of the mass into a characteristic redshift of formation. Unfortunately, this cannot be
done straightforwardly because our stellar ages are averages over all the stellar populations in a
galaxy, weighted by mass or by light. Thus they are very sensitive to the galaxy SFH and, on average,
closer to the last significant episode of star formation. Given the different timescales of star
formation expected in galaxies with different mass, it is reasonable to look at the distributions of mass
and metals against (mass-weighted) age for different stellar mass bins. This shows that most of the mass
and metals come from galaxies older than a characteristic age which becomes progressively younger from
massive to less massive galaxies. Considering only massive galaxies ($\rm >10^{11}M_\odot$) at least 50
percent of the mass and metals is in systems older than 8.5~Gyr, which would correspond to a redshift
$z>1.2$. This redshift is compatible with the redshift range $1<z<2$ at which the cosmic star formation
rate density starts to decline \citep[e.g.][]{hopkins06}. This is consistent with the evidence that the
bulk of star formation in massive galaxies has occurred at high redshifts, while low-mass galaxies are
still in a phase of active star formation \citep[e.g.][]{jarle2000,juneau05,bundy06}, although galaxies
appear to experience the same rate of decline in their global star formation from $z\sim$1 to the present
independently of mass \citep{xxz07}.

Comparison of the observed distributions of $\rho_Z$ and $\rho_\ast$ with those predicted by the
Millennium Simulation shows a good agreement in the distributions against stellar mass. However, 
an excess at low stellar masses (roughly below $\rm 3\times10^{10}M_\odot$) is visible, which has been
already pointed out in the luminosity or stellar mass functions \citep[see e.g.][]{croton06}. The
agreement between our observations and the models becomes worse when mass-weighted age and stellar
metallicity are considered. Although successful in reproducing the general trend of increasing both age
and metallicity with mass \citep[e.g.][]{delucia05}, the models predict narrower ranges in these physical
parameters than observed.
 
Our study provides a new determination of the total amount of metals locked up in stars today and allows
us for the first time to derive a quantitative and accurate description of the properties of the galaxies
hosting different fractions of the metals and of the stellar mass in the local Universe. Such
distributions represent important constraints on models of the cosmic star formation and chemical
enrichment histories. The detailed knowledge of the distribution of metals, coupled to the stellar mass
distribution, will allow a more direct comparison with predictions from semianalytic models of galaxy
formation and evolution. Ongoing large redshift surveys, like VVDS \citep{lefevre04b}, GOODS
\citep{goods,goods_spec}, DEEP2 \citep{deep2,deep2_spec}, zCOSMOS \citep{zcosmos}, will make it possible to
extend this kind of studies to redshifts as high as $z\sim1$ and thus, not only to build a more
consistent picture of the cosmic star formation history, but also to understand which are the galaxies
that most strongly contributed to its evolution since $z=1$.    

\section*{Acknowledgements}

We wish to thank the referee, Ivan Baldry, for a constructive report.
A.G. thanks Eric Bell for insightful comments, Fabio Fontanot for discussions on the MORGANA code,
Gabriella De Lucia for clarifications on the Millennium database, and Simon Driver for sharing results in
advance of publication. A.G. acknowledges support by the DFG's Emmy Noether Program. J.B. acknowledges
the receipt of FCT grant SFRH/BPD/14398/2003.
 
The Millennium Simulation databases used in this paper and the web application providing online access to
them were constructed as part of the activities of the German Astrophysical Virtual Observatory. They are
publicly accessible at http://www.mpa-garching.mpg.de/millennium.

Funding for the creation and distribution of the SDSS Archive has been provided by the Alfred P. Sloan
Foundation, the Participating Institutions, the National Aeronautics and Space Administration, the
National Science Foundation, the US Department of Energy, the Japanese Monbukagakusho, and the Max Planck
Society. The SDSS Web site is http://www.sdss.org/. The Participating Institutions are the University of
Chicago, Fermilab, the Institute for Advanced Study, the Japan Participation Group, the Johns Hopkins
University, the Max Planck Institute for Astronomy (MPIA), the Max Planck Institute for Astrophysics
(MPA), New Mexico State University, Princeton University, the United States Naval Observatory, and the
University of Washington.

\label{lastpage}
\end{document}